# Room-temperature antiskyrmions and sawtooth surface textures in a noncentrosymmetric magnet with $S_4$ symmetry

Kosuke Karube[1*], Licong Peng[1*], Jan Masell[1], Xiuzhen Yu[1], Fumitaka Kagawa[1,2], Yoshinori Tokura[1,2,3] & Yasujiro Taguchi[1]

1. RIKEN Center for Emergent Matter Science (CEMS), Wako 351-0198, Japan.

2. Department of Applied Physics, University of Tokyo, Bunkyo-ku 113-8656, Japan.

3. Tokyo College University of Tokyo, Bunkyo-ku 113-8656, Japan.

* These authors equally contributed

**Topological spin textures have attracted much attention both for fundamental physics and spintronics applications. Among them, antiskyrmions possess a unique spin configuration with Bloch-type and Néel-type domain walls due to anisotropic Dzyaloshinskii-Moriya interaction (DMI) in the noncentrosymmetric crystal structure. However, antiskyrmions have thus far only been observed in a few Heusler compounds with $D_{2d}$ symmetry. Here, we report a new material $Fe_{1.9}Ni_{0.9}Pd_{0.2}P$ in a different symmetry class ($S_4$ symmetry), where antiskyrmions exist over a wide temperature region including room temperature, and transform to skyrmions upon changing magnetic field and lamella thickness. The periodicity of magnetic textures greatly depends on crystal thickness, and domains with anisotropic sawtooth fractals are observed at the surface of thick crystals, which are attributed to the interplay between dipolar interaction and DMI as governed by**

**crystal symmetry. Our findings provide a new arena to study antiskyrmions, and should stimulate further research on topological spin textures and their applications.**

Magnetic skyrmions and antiskyrmions with vortex-like topological spin textures have recently attracted increasing attention as a source of various emergent phenomena and their potential applications to spintronics devices such as racetrack memory [1]. One of the microscopic origins for these topological textures is the competition between ferromagnetic exchange interaction and Dzyaloshinskii-Moriya interaction (DMI), the latter of which is derived from relativistic spin-orbit interaction in the absence of spatial inversion symmetry. Several types of DMI-based topological spin textures have been theoretically proposed [2, 3]. Skyrmions of both Bloch-type (Fig. 1a, b) and Néel-type (Fig. 1c, d) are characterized by an integer topological number [1]. A Bloch-type skyrmion is constructed by helical spin propagations (Bloch walls) with either of clockwise (CW) (Fig. 1a) or counterclockwise (CCW) (Fig. 1b) rotation, and has been observed in cubic chiral magnets belonging to $T$ and $O$ symmetry classes [4−7]. A Néel-type skyrmion is produced by cycloidal spin propagations (Néel walls) also with either of CW (Fig. 1c) or CCW (Fig. 1d) orientation, and has been observed in heterostructures with interfacial DMI [8−10] and bulk polar magnets with $C_{nv}$ symmetry [11−13].

An antiskyrmion, on the other hand, is characterized by a topological number with the opposite sign [1] and composed of spin spirals which cover all helicities (Fig 1e, f). This unique spin configuration is possible only when the DMI along two orthogonal axes have opposite signs, thus antiskyrmions are theoretically predicted to form in $D_{2d}$ and $S_4$ symmetry classes with a four-fold rotoinversion axis ($\bar{4}$) [2, 3, 14, 15]. Experimentally, antiskyrmions have been observed in inverse Heusler compounds with $D_{2d}$ symmetry (space group: $I\bar{4}2m$), such as Mn-deficient $Mn_{1.4}Pt_{0.9}Pd_{0.1}Sn$ (and $Mn_{1.4}PtSn$) [16] and stoichiometric $Mn_2Rh_{0.05}Ir_{0.05}Sn$ [17]. In contrast to extensive studies of skyrmions in various magnets, previous experimental studies for antiskyrmions have been confined to these $D_{2d}$ Heusler alloys, and no other materials have been found. Here, we report our discovery of a new material $Fe_{1.9}Ni_{0.9}Pd_{0.2}P$ with $S_4$ symmetry, in which antiskyrmions and skyrmions are observed above room temperature by using Lorentz transmission electron microscopy (LTEM). In conjunction with magnetic force microscopy (MFM), we also describe the evolution of bulk and surface magnetic textures reflecting the anisotropic DMI and underlying $S_4$-type lattice symmetry as the crystal thickness is increased.

**Basic magnetic properties of $Fe_{1.9}Ni_{0.9}Pd_{0.2}P$**

Our target compound $M_3P$ ($M$: transition metal) crystalizes in a noncentrosymmetric tetragonal structure with the space group of $I\bar{4}$ in the $S_4$ symmetry class (Fig. 1g). The structural symmetry is characterized by only $\bar{4}$ around the [001] axis, and thus lower than $D_{2d}$. Among $M_3P$ compounds, $Fe_3P$ is a ferromagnet with a high Curie temperature $T_c \sim 700$ K and saturation moment $M_s \sim 1.84$ $\mu_B$/Fe, while $Ni_3P$ is a Pauli paramagnet without showing magnetic orders [18, 19]. In their solid solutions (Fe, Ni)$_3$P called "schreibersite", $T_c$ and $M_s$ linearly decrease as the Ni concentration is increased [18, 19].

We obtained high-quality bulk single crystals of $Fe_{1.9}Ni_{0.9}Pd_{0.2}P$ (see Supplementary Note 1, Tables 1, 2 and Fig. 1 for sample characterization). The Pd doping to the schreibersite was anticipated to enhance the spin-orbit coupling and hence DMI, and it also turned out to change magnetic anisotropy from in-plane to uniaxial one (see Supplementary Note 2 and Fig. 2 for magnetization data in $Fe_{1.6}Ni_{1.4}P$). Temperature dependence of the magnetization at 0.01 T (Fig. 1h) shows a ferromagnetic transition at $T_c \sim 400$ K. As presented in Fig. 1i, the magnetization at 2 K reaches the saturation value of $M_s \sim 3.5$ $\mu_B$/f.u. at 0.35 T for [001] while at 0.64 T for [110], clearly indicating easy-axis anisotropy. There is no hysteresis for both directions. We derived the effective uniaxial anisotropy energy $K_u \sim 106$ kJ/m$^3$ from the area enclosed between the [001] and [110] magnetization curves for $H > 0$. The quality factor $Q$, defined as the ratio between uniaxial anisotropy and magnetostatic energies $K_u/(\mu_0 M_s^2/2)$, is obtained to be $Q \sim 0.4$ at 2 K (0.28 at 300 K). Therefore, Fe$_{1.9}$Ni$_{0.9}$Pd$_{0.2}$P is a soft magnet with relatively strong dipolar interaction and weak uniaxial anisotropy, both of which are important for the formation of antiskyrmions [15, 20], similarly to the Heusler alloys [16, 21].

**LTEM observation of antiskyrmions and skyrmions**

For the real-space observation of magnetic structures, we performed LTEM measurements on (001) thin plates. Figure 2 summarizes the LTEM results at 295 K for a fixed sample thickness $t \sim 130$ nm. The magnetic structures strongly depend on measurement protocols, especially tilting of the sample plate under magnetic fields (see details in Supplementary Note 3 and Fig. 3). At zero field, helical stripe pattern with a periodicity of $\lambda \sim 280$ nm is observed (Fig. 2b). Here, helical propagation vectors (***q***) are pinned almost along the [110] and [$\bar{1}$10] axes, and the helicity of Bloch-type domain wall is reversed between them. This result evidences the presence of anisotropic DMI as expected for the $S_4$ crystal symmetry. Since the [110] and [$\bar{1}$10] axes are not experimentally distinguished, we use [110] for the CW rotation axis and [$\bar{1}$10] for the CCW one in this paper.

Under magnetic fields perpendicular to the plate, antiskyrmions sparsely form (Supplementary Fig. 3a). When the sample plate is tilted by 12° and then a magnetic field is applied, a large number of magnetic bubbles with zero topological number, termed here non-topological (NT-) bubbles as in Ref. [22], are created above 350 mT (Supplementary Fig. 3b). Next, while the plate is tilted back toward the initial orientation at 350 mT, the NT-bubbles deform to the bullet-like shape (Fig. 2c), and eventually convert to antiskyrmions. As shown in Fig. 2d, the antiskyrmions are square-shaped with the longer Bloch-wall parts along the [110] and [$\bar{1}$10] axes, and form a lattice state. As the field is further increased, the antiskyrmions persist up to 470 mT and finally transform to Bloch-type skyrmions at 500 mT via a coexistence state with NT-bubbles around 480−490 mT (see Supplementary Video 1 for the details of the transformation). As shown in Fig. 2e, CW and CCW skyrmions coexist (in almost equal proportion in a larger area) and are elliptically elongated along the [$\bar{1}$10] and [110] directions, respectively. The density of NT-bubbles, antiskyrmions, and skyrmions is plotted against the magnetic field in Fig. 2f, clearly showing the systematic transformation among the topologically distinct three magnetic objects through the creation and annihilation of Bloch line pair. These LTEM measurements are performed at different temperatures, and similar field-induced transformations from antiskyrmions to skyrmions are observed at almost all measured temperatures below $T_c$ down to 100 K, as summarized in Fig. 2g.

Similar square antiskyrmions and elliptical Bloch-type skyrmions with different helicities have been previously reported in $Mn_{1.4}Pt_{0.9}Pd_{0.1}Sn$ [22, 23], and explained in terms of the competition between anisotropic DMI and dipolar interaction. In general, dipolar interactions prefer Bloch walls to Néel walls to minimize magnetic volume charges. Therefore, the Bloch-wall parts along [110] and [$\bar{1}$10] directions in antiskyrmions (red and blue lines in Fig. 1e) expand while the Néel-wall parts along [100] and [010] directions (green and yellow lines in Fig. 1e) shrink, resulting in square-shaped antiskyrmions [15, 22, 23]. The elliptical shapes of skyrmions whose elongated directions are rotated by 90° for the different helicities are attributed to the anisotropic DMI: Along [110] and [$\bar{1}$10] directions, CW and CCW Bloch walls are respectively favored by the anisotropic DMI, therefore the CW and CCW skyrmions are elliptically elongated along different directions [22, 23]. Although the shape of (anti)skyrmions is similar to those in the Heusler alloys, their orientations with respect to the tetragonal crystalline axes are different between the two systems because the Bloch wall directions are fixed to be $q \parallel$ [110] and [$\bar{1}$10] in $Fe_{1.9}Ni_{0.9}Pd_{0.2}P$ while they are $q \parallel$ [100] and [010] in $Mn_{1.4}Pt_{0.9}Pd_{0.1}Sn$ [16, 22, 23]. Importantly, the topological transition between the antiskyrmion and the skyrmion phases in the present compound is controllable by the magnetic field over the whole temperature region below $T_c$ including room temperature, making a striking contrast with the case of $Mn_{1.4}Pt_{0.9}Pd_{0.1}Sn$, where the two phases are separated by temperature around 265 K [22, 23].

As the magnetic textures are strongly influenced by dipolar interactions, we carried out LTEM measurements for a lamella with various thicknesses ranging from 50 nm to 210 nm. The LTEM images at 295 K and 400 mT after a similar tilting procedure are presented for $t \sim$ 50 nm, 100 nm, and 210 nm in Fig. 3a-c, respectively. At the thinnest area with $t \sim$ 50 nm, elliptical skyrmions are observed (Fig. 3a). On the other hand, square antiskyrmions are discerned at thick areas above $t \sim$ 130 nm up to 210 nm (Fig. 3c). At an intermediate thickness of $t \sim$ 100 nm, elliptical skyrmions and square antiskyrmions coexist (Fig. 3b). Clearly, the size of the (anti)skyrmions becomes larger as the lamella thickness increases, similarly to the increase in helical periodicity at zero field (Supplementary Fig. 5). The thickness-dependent transformation between the two topological textures is also understood in terms of the competition between dipolar interaction and DMI. In a sufficiently thin plate with a large surface-to-volume ratio, to reduce the magnetostatic energy, Bloch-type skyrmions without Néel-like configuration are favored. On the other hand, in a thick sample with a large volume, dipolar interactions are less important than DMI, and thus antiskyrmions are preferred.

**MFM observation of sawtooth domain patterns**

Next, we performed MFM measurements to observe surface magnetic textures, such as helical stripes and antiskyrmions, for thicker samples ranging from $t \sim 0.52$ $\mu$m up to 240 $\mu$m. Figures 4b-f show MFM images for different thickness, taken at 295 K and 0 mT after application of magnetic fields with tilting the samples, as illustrated in Fig. 4a. For $t \sim 0.52$ $\mu$m (Fig. 4b), square objects are clearly observed in addition to stripe domains. Although MFM detects the out-of-plate component of the stray magnetic field from the domains on the surface, the MFM image bears a strong resemblance with the LTEM images in zero field (Supplementary Fig. 3d), hence we can safely conclude that the square objects correspond to the antiskyrmions persisting at zero field. As the thickness is increased, the stripe periodicity and the antiskyrmion size further increase, and their domain structures become more complex. The domain boundaries exhibit clear undulations for $t \sim 2.2$ $\mu$m (Fig. 4c), which turns into prominent sawtooth waves for $t \sim$ 5.3 $\mu$m (Fig. 4d), where core structures with reversed magnetization are also observed as indicated with a dotted circle. These complex domain textures are further branched and characterized as fractals, i.e., self-similar objects in multiple length scales, for $t \sim 50$ $\mu$m (Fig. 4e and f). Note that the square objects enclosed by the vertical and horizontal sawtooth stripes with $q \parallel [110]$ and $q \parallel [\bar{1}10]$ are characterized by $\bar{4}$ symmetry as schematically illustrated in Fig. 4g, which is identical to the symmetry of the underlying crystal lattice.

Here, we discuss the observed complex domain pattern. In general, magnetic domains are formed so that the magnetostatic energy due to magnetic surface charges is reduced at the cost of increased domain wall energy [24−26]. It has been known since 1960's that as the crystal thickness increases, magnetic domains at the surface become more complex, and show corrugations and even fractalizations to further screen the surface magnetic charges on larger scales [26−30]. Such wavy or branched flower-like domain textures have been observed in a broad range of uniaxial ferromagnets such as Co [26, 28], Nd-Fe-B alloys [26, 30, 31], hexagonal ferrites [28, 32] and garnet ferrites [29]. In spite of such a long history of domain wall physics over 60 years, the anisotropic sawtooth pattern governed by $\bar{4}$ symmetry as observed in this antiskyrmion material has never been reported so far, and thus must be exclusively attributed to the anisotropic DMI in the $S_4$ crystal lattice.

To understand the role of the DMI for the formation of the sawtooth pattern, we have performed micromagnetic simulations for a three-dimensional system using the program MuMax3 [33] (see Methods and Supplementary Note 9 for the details). In the simulation, we incorporate the exchange stiffness, the uniaxial anisotropy energy and the demagnetization energy using the experimentally evaluated values at 300 K, and introduce the anisotropic DMI as a small perturbation. We found that slightly asymmetric wave appears near the surface without DMI, but the DMI clearly enhances the asymmetric feature of the wave and leads to the sawtooth pattern as presented in Extended Data Fig. 1, which is similar to the MFM result for the crystal with $t \sim 5.3 \ \mu$m. The sawtooth pattern can be well understood in terms of the helicity of the domain walls uniquely determined by the anisotropic DMI in the bulk, and the additional modification caused by the competition between the domain wall energy and demagnetization energy near the surface (see Supplementary Note 10 and Fig. 7 for the detailed discussion). The micromagnetic simulation indicates that the sawtooth pattern reverses for the top surface and the bottom surface when viewed from the same direction, which is attributed to the opposite direction of the stray fields for the two surfaces. This is confirmed by the additional MFM measurement on the opposite side of the crystal (Supplementary Fig. 7).

**Thickness dependence of magnetic periodicity**

In Fig. 4h, the magnetic periodicity $\lambda$ determined from the LTEM and MFM measurements is plotted against the thickness $t$ in a wide range of 50 nm − 240 $\mu$m. $\lambda$ shows a dramatic increase from the order of 100 nm up to the order of 10 $\mu$m as $t$ is increased. Such significant dependence of $\lambda$ on $t$ up to micrometer-scale is a typical consequence of dipolar interaction as commonly observed in the conventional uniaxial ferromagnets [28−30, 32], and also in the antiskyrmion-host Heusler compound as recently reported [34]. The observed $\lambda(t)$ can be fitted to a power law as $\lambda \sim t^{0.46}$ below $t \sim 10$ $\mu$m, and $\lambda \sim t^{0.65}$ above $t \sim 10$ $\mu$m. The obtained power-law exponents are in accordance with those in theoretical models for uniaxial ferromagnets. As described in the Supplementary Note 8 in more detail, the well-known Kittel's model for sufficiently thin films gives $\lambda \sim t^{1/2}$ [24], and other theories applicable to thick crystals with wavy or spiky domain structures near the surface predict $\lambda \sim t^{0.6}$ [27] or $\lambda \sim t^{2/3}$ [26, 28]. While the DMI is not considered in these theories, the overall feature of the thickness evolution as governed by the dipolar interaction can be captured. On the basis of these theoretical models, we calculate the critical thickness $t_c \sim 0.4$ $\mu$m at 300 K, above which undulations of surface domain walls take place, in good agreement with the MFM data. This value is very small as compared with those previously reported for uniaxial ferromagnets (typically $t_c \sim 10$ $\mu$m [30]) due to the relatively weak uniaxial anisotropy ($Q < 1$) in the present material. In the recent MFM study for $Mn_{1.4}PtSn$ [34], surface patterns of magnetic stripes and antiskyrmions remain simple and clear undulations of domain walls are absent up to $t \sim 4$ $\mu$m, probably because the $Q$ value is relatively large in the Heusler compounds. Associated with the weak magnetic anisotropy, another vortex-like structure within (anti)skyrmions is observed by LTEM (Figs. 2, 3), which is also absent in the Heusler alloys. The magnetic softness of the present material as compared with the Heusler compounds is another interesting point in these two different antiskyrmion systems.

**Summary and outlook**

The present work demonstrates that the Pd-doped schreibersite, $Fe_{1.9}Ni_{0.9}Pd_{0.2}P$, is a new class of room-temperature antiskyrmion material with $S_4$ symmetry, and that antiskyrmions and skyrmions can be easily interconverted by changing magnetic field and crystal thickness over a wide temperature region. The interplay between anisotropic DMI and long-range dipolar interactions provides a rich variety of magnetic textures with variable size, shape and topology, including surface sawtooth fractals as dominated by the crystal symmetry. These new findings accelerate the studies of antiskyrmions and related topological spin textures, and open an entirely new chapter for domain wall physics, where the DMI-induced asymmetric features have been overlooked for over 60 years. Furthermore, these topological and anisotropic spin textures with versatile tunability at room temperature are promising for spintronics applications.

**Methods**

**Sample preparation**

Single-crystalline bulk samples of $Fe_{1.6}Ni_{1.4}P$ and $Fe_{1.9}Ni_{0.9}Pd_{0.2}P$ were synthesized by a self-flux method from pure Fe, Ni and Pd metals and red phosphorous sealed in an evacuated quartz tube. The target phase of tetragonal $M_3P$ was isolated from the ingot. Phase purity with the $M_3P$ structure was confirmed by powder X-ray diffraction with Cu K$\alpha$ radiation as detailed in Supplementary Fig. 1. Crystal orientations were checked by X-ray Laue diffraction method. Chemical compositions were examined by a scanning electron microscope (SEM) equipped with an energy dispersive X-ray (EDX) analyzer (Supplementary Tables 1, 2).

**Magnetization measurement**

For the magnetization measurement, a single-crystalline bulk piece was cut to a rectangular shape (0.5 mm × 0.5 mm × 1.0 mm) with almost the same surface areas of (001) and (110) planes. Magnetization measurements were performed by a superconducting quantum interference device magnetometer (MPMS3, Quantum Design) equipped with an oven option.

**LTEM measurement**

For the Lorentz transmission electron microscopy (LTEM) measurements, (001) thin plates with various thickness ($t \sim$ 50 nm, 70 nm, 100 nm, 130 nm, 160 nm and 210 nm) were thinned from the bulk crystals by a focused ion beam (FIB) system (NB5000, Hitachi). LTEM measurements were performed with a transmission electron microscope (JEM-2100F, JEOL) equipped with a double-tilt liquid-nitrogen holder (Gatan 636) and a double-tilt heating holder (Protochips: Fusion select). External magnetic fields applied to the (001) plates were obtained by tuning the objective lens current of JEM-2100F, which are parallel to the incident electron beam. An in-plane component of the field was applied by tilting the plate as illustrated in Fig. 2a. LTEM defocus images with bright and dark contrasts reflect the in-plane component of magnetic induction fields averaged over the plate thickness. The lateral magnetic induction field distribution was obtained by the transport-of-intensity equation (TIE) analysis [35] of the over-focus and under-focus LTEM images and displayed by color coding.

**DPC-STEM measurement**

For additional information to support the TIE data, we performed the differential phase contrast (DPC) imaging in scanning transmission electron microscope (STEM) (JEM-2100F, JEOL) equipped with a segmented annular all field (SAAF) detector. The DPC-STEM observations for the same plate with that for LTEM observations, were performed at room temperature and zero field in a Low-Mag mode where the objective-lens current is zero. The software package qDPC (HREM Co.) was used to analyze the raw data obtained by DPC-STEM [43].

**MFM measurement**

For the magnetic force microscopy (MFM) measurements, a staircase-shaped plate with (001) surfaces and various thickness ($t \sim$ 520 nm, 2.2 $\mu$m, 5.3 $\mu$m and 10.3 $\mu$m) was prepared from a bulk piece by FIB. In addition, bulk crystals with thickness of $t \sim 50$ $\mu$m, 140 $\mu$m and 240 $\mu$m were prepared, and their (001) surfaces were treated by chemical mechanical polishing (CMP) with colloidal silica. For the purpose of MFM observations of metastable antiskyrmions at zero magnetic field, a magnetic field was applied perpendicular to the (001) plane in Physical Properties Measurement System (PPMS, Quantum Design), and an in-plane component of the field was applied by tilting the plate as depicted in Fig. 4a. MFM measurements were performed at room temperature (295 K ±3 K) and zero field with a commercial scanning probe microscope (MFP-3D, Asylum Research). We used the MFM cantilever (MFMR, Nano World) with a resonance frequency of ~ 72 kHz and Co coating on the whole tip to keep the magnetization of the tip along the direction perpendicular to the sample surface. We used the two-pass technique: mapping a surface topography by tapping the sample in the first pass, and mapping an MFM image while lifting the cantilever from the sample (distance 30 nm − 50 nm) in the second pass. For the analysis, the flattening procedure was used. The color of the MFM images shows the phase shift of the oscillating cantilever, typically ranging from +2º (bright yellow) to −2º (dark purple) in this work, whose sign and amplitude correspond to those of the second-derivative of the stray magnetic field mainly produced by the magnetization perpendicular to the sample plate.

**Micromagnetic simulation**

For the micromagnetic simulation of the sawtooth domain pattern, we used the well-established GPU-accelerated program MuMax3 [33] and personalized the code to incorporate the asymmetric DMI and an improved numerical accuracy. The energy functional is given by

$$E[m] = \int_V A\,(\nabla \boldsymbol{m})^2 + D\,[\boldsymbol{m}\cdot(\hat{x}\times\partial_x\boldsymbol{m}) - \boldsymbol{m}\cdot(\hat{y}\times\partial_y\boldsymbol{m})] - K_{\mathrm{u}}\,(\boldsymbol{m}\cdot\hat{z})^2 - \frac{M_{\mathrm{s}}}{2}\boldsymbol{m}\cdot\boldsymbol{B}_{\mathrm{d}}\,dV$$

where $A$, $D$, $K_{\mathrm{u}}$, and $M_{\mathrm{s}}$ correspond to the exchange stiffness, the DMI constant, the uniaxial anisotropy energy constant, and the saturation magnetization, respectively. $\boldsymbol{B}_{\mathrm{d}}$ represents the demagnetizing field. For the simulation, we use the experimentally evaluated parameters as $A = 8.1\text{ pJ/m}$, $K_{\mathrm{u}} = 31\text{ kJ/m}^3$, and $M_{\mathrm{s}} = 417\text{ kA/m}$, and treat the DMI as a small perturbation with various values $D = 0, 0.05, 0.1, 0.2\text{ mJ/m}^2$ (see Supplementary Note 9 for more details). The simulated sample size of 1.6 μm × 0.8 μm × 5.3 μm with periodic boundary conditions in the x-y-plane is chosen according to the experimental observations and is discretized to 512 × 256 × 256 cuboids. The simulated sawtooth pattern for $D = 0.2\text{ mJ/m}^2$ is presented in Extended Data Fig. 1.

**References**

1. Nagaosa, N. & Tokura, Y. Topological properties and dynamics of magnetic skyrmions. *Nat. Nanotech.* **8**, 899−911 (2013).

2. Bogdanov, A. N. & Yablonskii, D. A. Thermodynamically stable "vortices" in magnetically ordered crystals. The mixed state of magnets. *Sov. Phys. JETP* **68**, 101−103 (1989).

3. Leonov, A. O. *et al*. The properties of isolated chiral skyrmions in thin magnetic films. *New J. Phys.* **18**, 065003 (2016).

4. Mühlbauer, S. *et al*. Skyrmion lattice in a chiral magnet. *Science* **323**, 915−919 (2009).

5. Yu, X. Z. *et al*. Real-space observation of a two-dimensional skyrmion crystal. *Nature* **465**, 901−904 (2010).

6. Seki, S., Yu, X. Z., Ishiwata, S. & Tokura, Y. Observation of skyrmions in a multiferroic material. *Science* **336**, 198−201 (2012).

7. Tokunaga, Y. *et al*. A new class of chiral materials hosting magnetic skyrmions beyond room temperature. *Nat. Commun.* **6**, 7638−7644 (2015).

8. Heinze, S. *et al*. Spontaneous atomic-scale magnetic skyrmion lattice in two dimensions. *Nat. Phys.* **7**, 713−718 (2011).

9. Woo, S. *et al*. Observation of room-temperature magnetic skyrmions and their current-driven dynamics in ultrathin metallic ferromagnets. *Nat. Mater.* **15**, 501–506 (2016).

10. Pollard, S. D. *et al*. Observation of stable Néel skyrmions in cobalt/palladium multilayers with Lorentz transmission electron microscopy. *Nat. Commun.* **8**, 14761 (2017).

11. Kézsmárki, I. *et al*. Néel-type skyrmion lattice with confined orientation in the polar magnetic semiconductor $GaV_4S_8$. *Nat. Mater.* **14**, 1116−1122 (2015).

12. Kurumaji, T. *et al*. Néel-type skyrmion lattice in the tetragonal polar magnet $VOSe_2O_5$. *Phys. Rev. Lett.* **119**, 237201 (2017).

13. Srivastava, A. K. *et al*. Observation of Robust Néel Skyrmions in Metallic PtMnGa. *Adv. Mater.* **32**, 1904327 (2020).

14. Hoffmann, M. *et al*. Antiskyrmions stabilized at interfaces by anisotropic Dzyaloshinskii–Moriya interactions. *Nat. Commun.* **8**, 308 (2017).

15. Camosi, L., Rougemaille, N., Fruchart, O., Vogel, J. & Rohart, S. Micromagnetics of antiskyrmions in ultrathin films. *Phys. Rev. B* **97**, 134404 (2018).

16. Nayak, A. K. *et al*. Magnetic antiskyrmions above room temperature in tetragonal Heusler materials. *Nature* **548**, 561–566 (2017).

17. Jena, J. *et al*. Observation of Magnetic Antiskyrmions in the Low Magnetization Ferrimagnet $Mn_2Rh_{0.95}Ir_{0.05}Sn$. *Nano Lett.* **20**, 59–65 (2020).

18. Gambino, R. J., McGuire, T. R. & Nakamura, Y. Magnetic Properties of the Iron-Group Metal Phosphides. *J. Appl. Phys.* **38**, 1253−1255 (1967).

19. Goto, M., Tange, H., Tokunaga, T., Fujii, H. & Okamoto, T. Magnetic Properties of the $(Fe_{1-x}M_x)_3P$ Compounds. *Jpn. J. Appl. Phys.* **16**, 2175−2179 (1977).

20. Koshibae, W. & Nagaosa, N. Theory of antiskyrmions in magnets. *Nat. Commun.* **7**, 10542 (2016).

21. Vir, P. *et al*. Tetragonal Superstructure of the Antiskyrmion Hosting Heusler Compound $Mn_{1.4}PtSn$. *Chem. Mater.* **31**, 5876−5880 (2019).

22. Peng, L. C. *et al*. Controlled transformation of skyrmions and antiskyrmions in a non-centrosymmetric magnet. *Nat. Nanotech.* **15**, 181–186 (2020).

23. Jena, J. *et al*. Elliptical Bloch skyrmion chiral twins in an antiskyrmion system. *Nat. Commun.* **11**, 1115 (2020).

24. Kittel, C. Theory of the Structure of Ferromagnetic Domains in Films and Small Particles. *Phys. Rev.* **70**, 965–971 (1946).

25. Malozemoff, A. P. & Slonczewski, J. C. Magnetic Domain Walls in Bubble Materials (Academic Press, New York, 1979).

26. Hubert, A. & Schäfer, R. Magnetic Domains: The Analysis of Magnetic Microstructures (Springer, Berlin, 1998).

27. Szymczak, R. A Modification of the Kittel Open Structure. *J. Appl. Phys.* **39**, 875–876 (1968).

28. Kaczér, J. On the domain structure of uniaxial ferromagnets. *Sov. Phys. JETP* **19**, 1204–1208 (1964).

29. Gemperle, R., Murtinová, L. & Kamberský, V. Experimental Verification of Theoretical Relations for the Domain Structure of Uniaxial Ferromagnets. *Phys. Stat. Sol. A* **158**, 229 (1996).

30. Szmaja, W. Investigations of the domain structure of anisotropic sintered Nd–Fe–B-based permanent magnets. *J. Mag. Mag. Mater.* **301** 546–561 (2006).

31. Kreyssig, A. *et al*. Probing Fractal Magnetic Domains on Multiple Length Scales in $Nd_2Fe_{14}B$. *Phys. Rev. Lett.* **102**, 047204 (2009).

32. Jalli, J. *et al*. MFM studies of magnetic domain patterns in bulk barium ferrite ($BaFe_{12}O_{19}$) single crystals. *J. Mag. Mag. Mater.* **323**, 2627–2631 (2011).

33. Vansteenkiste, A. *et al*. The design and verification of MuMax3. *AIP Adv.* **4**, 107133 (2014).

34. Ma, T. *et al*. Tunable Magnetic Antiskyrmion Size and Helical Period from Nanometers to Micrometers in a $D_{2d}$ Heusler Compound. *Adv. Mater.* **32**, 2002043 (2020).

35. Ishizuka, K. & Allman, B. Phase measurement of atomic resolution image using transport of intensity equation. *J. Electron Microsc.* **54**, 191–197 (2005).

36. Chapman, J. N., Batson, P. E., Waddell, E. M. & Ferrier, R. P. The direct determination of magnetic domain wall profiles by differential phase contrast electron microscopy. *Ultramicroscopy* **3**, 203–214 (1978).

37. Sandweg, C. W. *et al*. Direct observation of domain wall structures in curved permalloy wires containing an antinotch. *J. Appl. Phys.* **103**, 093906 (2008).

38. McGrouther, D. *et al*. Internal structure of hexagonal skyrmion lattices in cubic helimagnets. *New J. Phys.* **18**, 095004 (2016).

39. Shibata, N. *et al*. Direct Visualization of Local Electromagnetic Field Structures by Scanning Transmission Electron Microscopy. *Acc. Chem. Res.* **50**, 1502−1512 (2017).

40. Matsumoto, T., So, Y. G., Kohno, Y., Ikuhara, Y. & Shibata, N. Stable Magnetic Skyrmion States at Room Temperature Confined to Corrals of Artificial Surface Pits Fabricated by a Focused Electron Beam. *Nano Lett.* **18**, 754–762 (2018).

41. Pöllath, S. *et al*. Spin structure relation to phase contrast imaging of isolated magnetic Bloch and Néel skyrmions. *Ultramicroscopy* **212**, 112973 (2020).

42. Yasin, F. S. *et al*. Bloch lines constituting antiskyrmions captured via differential phase contrast. *Adv. Mater.* **32**, 2004206 (2020).

43. Ishizuka, A., Oka, M., Seki, T., Shibata, N. & Ishizuka, K. Boundary-artifact-free determination of potential distribution from differential phase contrast signals. *Microscopy* **66**, 397 (2017).

**Acknowledgments**

We are grateful to N. Nagaosa, W. Koshibae, Y. Tokunaga and T. Arima for fruitful discussions. We also thank F. S. Yasin for technical support for the DPC-STEM measurement and K. Nakajima for technical support for preparation of the FIB sample. This work was supported by JSPS Grant-in-Aids for Scientific Research (Grant No. 17K18355, No. 18H05225, No. 19H00660 and No. 20K15164), JST CREST (Grant No. JPMJCR1874, No. JPMJCR20T1), and the Humboldt/JSPS International Research Fellow program (Grant No. 19F19815).

**Author contributions**

K.K., X.Y., Y. Tokura and Y. Taguchi jointly conceived the project. K.K. synthesized bulk crystals and performed magnetization measurements. L.P. fabricated FIB samples and performed LTEM and DPC-STEM measurements. MFM measurement was performed by K.K. with the support of F.K. J.M. theoretically considered the experimental results and performed micromagnetic simulations. The results were discussed and interpreted by all the authors.

**Additional information**

Correspondence and requests for materials should be addressed to K.K. (kosuke.karube@riken.jp) or Y. Taguchi (y-taguchi@riken.jp).

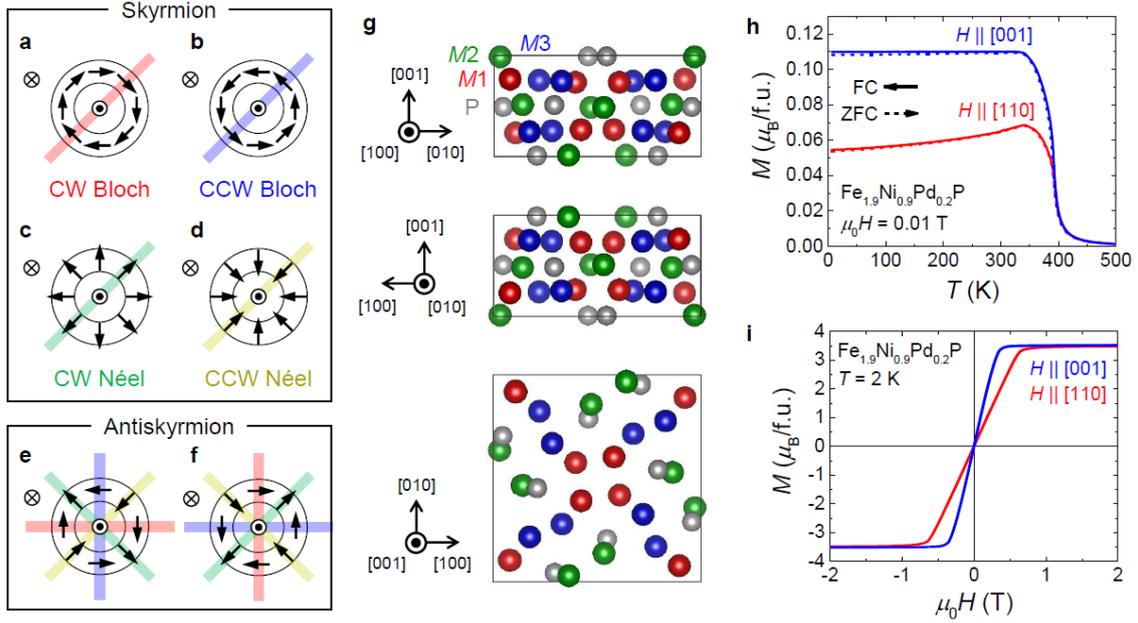

**Figure 1 | Antiskyrmions and basic properties of Fe$_{1.9}$Ni$_{0.9}$Pd$_{0.2}$P. a-f,** Schematics of (a, b) Bloch-type skyrmions, (c, d) Néel-type skyrmions and (e, f) antiskyrmions. Black arrows denote the directions of in-plane magnetic moments. Along red and blue lines, magnetic structures are clockwise (CW) and counterclockwise (CCW) Bloch walls, respectively, while along green and yellow lines, CW and CCW Néel walls, respectively. For antiskyrmions, the four different walls alternately appear every 45° along the azimuthal angle direction in the order of CW Bloch (red) → CW Néel (green) → CCW Bloch (blue) → CCW Néel (yellow). **g,** Schematic of noncentrosymmetric tetragonal crystal structure of $M_3$P ($M$: transition metal) with space group of $I\bar{4}$ ($S_4$ symmetry) as viewed along the [100], [010] and [001] axes from the top to the bottom, respectively. There are three inequivalent crystallographic $M$ sites denoted as $M$1 (red), $M$2 (green) and $M$3 (blue). As discerned in these schematics, two-fold rotation symmetry around [100] and [010] axes and mirror symmetry with respect to (110) and ($\bar{1}$10) planes are absent in the $S_4$ symmetry, resulting in a lower symmetry than $D_{2d}$. According to the sample characterization as detailed in Supplementary information (Note 1, Tables 1, 2 and Fig.

1), our target compound $Fe_{1.9}Ni_{0.9}Pd_{0.2}P$ crystallizes in the $M_3P$ structure while there is random site disorder among the three transition metals at $M$ sites. However, the overall crystal symmetry in this mixed crystal is preserved in average, and the impact of disorder on the magnetic properties of this itinerant material is minimal. **h**, Temperature ($T$) dependence of magnetization ($M$) under the magnetic field $\mu_0H = 0.01$ T in a bulk single crystal of $Fe_{1.9}Ni_{0.9}Pd_{0.2}P$. Magnetizations for $H \parallel [001]$ and $H \parallel [110]$ are presented with blue and red lines, respectively. The data taken during a field cooling (FC) and a field warming after a zero-field cooling (ZFC) are denoted with solid and dotted lines, respectively. **i**, Magnetic field dependence of magnetization at 2 K for $H \parallel [001]$ (blue line) and $H \parallel [110]$ (red line).

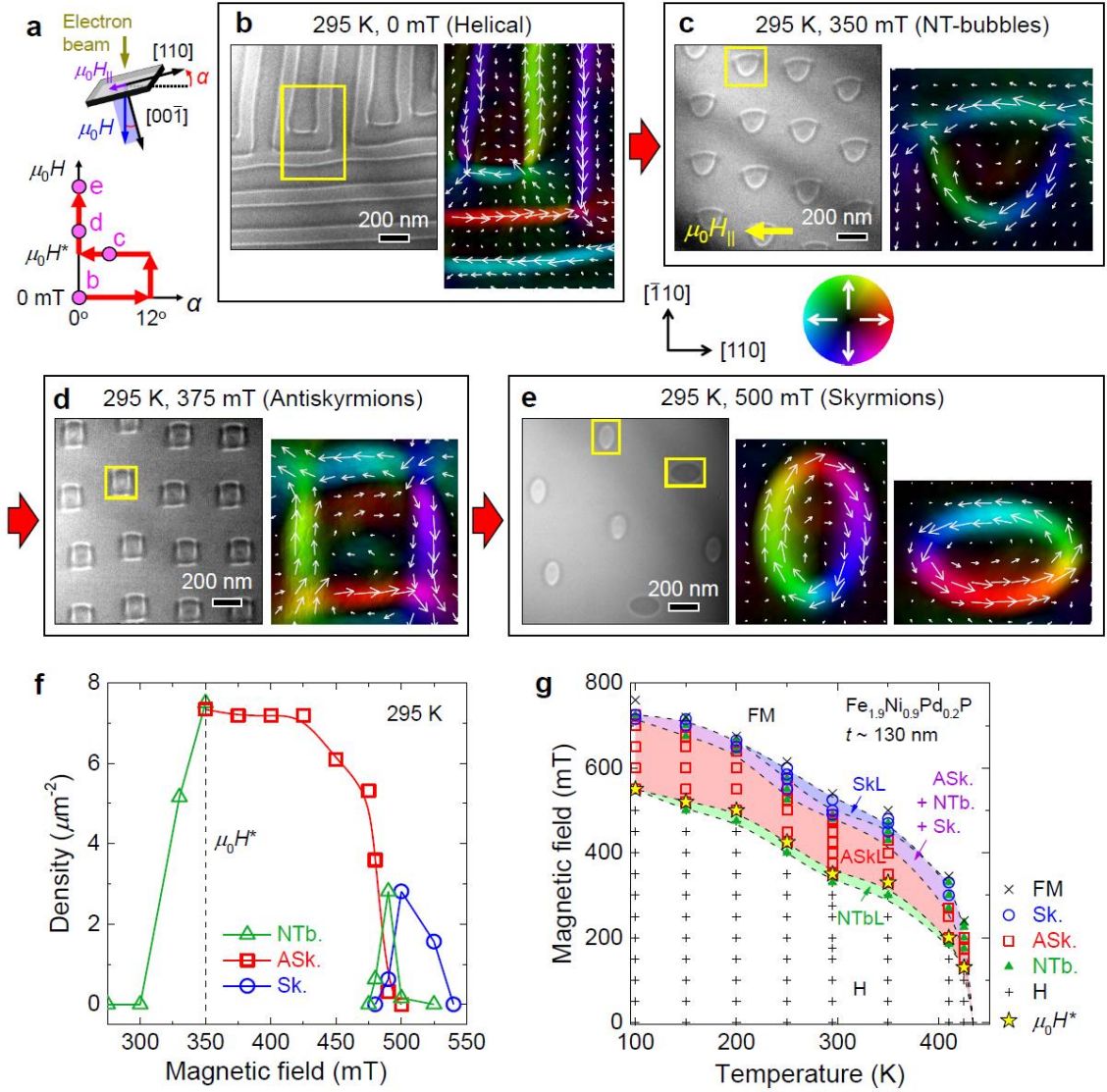

**Figure 2 | Magnetic textures in a lamella with a thickness $t \sim 130$ nm. a**, Schematics of the experimental configuration and the measurement process for panels b-g: (i) Tilting the sample plate by $\alpha \sim 12°$ at zero field → (ii) Increasing the magnetic field → (iii) Tilting the sample plate back to $\alpha \sim 0°$ at $\mu_0 H^*$ (350 mT for 295 K) → (iv) Increasing the field. Here, the magnetic field direction is antiparallel to [001], and the sample plate is tilted by the angle $\alpha$ from the plane perpendicular to the electron beam and the field so that the in-plane field direction is antiparallel to [110]. The magnetic field $\mu_0 H^*$ is chosen as the field where non-topological (NT-) bubbles are first observed at respective

temperatures. **b-e**, Under-focus LTEM images at 295 K and at several fields: (b) 0 mT, (c) 350 mT (with tilt of $\alpha \sim 5°$), (d) 375 mT ($\alpha \sim 0°$) and (e) 500 mT ($\alpha \sim 0°$), taken in the process shown in panel a. The defocus value for all the LTEM images is 200 $\mu$m. The direction of the in-plane bias field ($\mu_0 H_\parallel$) caused by the tilt is indicated with the yellow allow in the panel c. Color mapping in each panel shows the in-plane magnetic induction fields (*B*-fields) deduced by transport-of-intensity equation (TIE) analyses [35] of defocused LTEM images for the area indicated by a yellow rectangle. We also performed the differential phase contrast (DPC) imaging at scanning transmission electron microscopy (STEM) mode, which has been utilized to observe magnetic domain walls [36, 37], skyrmions [38−41] and antiskyrmions [42]. As detailed in Supplementary Note 5 and Fig. 4, the obtained *B*-field of antiskyrmions by DPC-STEM is the same with the TIE results. **f**, Magnetic field dependence of the density of NT-bubbles (NTb.), antiskyrmions (ASk.) and skyrmions (Sk.) counted in a fixed area of the LTEM images (~ 6.4 $\mu$m$^2$) in the process shown in the panel a (295 K, $\mu_0 H^* = 350$ mT). **g**, Magnetic phase diagram on the plane of temperature and magnetic field determined by the LTEM measurement in the field increasing process shown in panel a. The observed magnetic textures at each temperature and magnetic-field point are denoted with different symbols and colors as defined at the right side of the panel. An asterisk at each temperature indicates the magnetic field $\mu_0 H^*$ where the sample plate is tilted back to 0°. In the phase diagram we use the following abbreviations: H (helical phase), NTbL (NT-bubble lattice), ASkL (antiskyrmion lattice), SkL (skyrmion lattice), FM (field-induced ferromagnetic phase). The region where two or all of the different magnetic objects (antiskyrmions, NT-bubbles and skyrmions) coexist is shown with purple color. LTEM images and phase diagrams in different measurement protocols are presented in Supplementary Fig. 3.

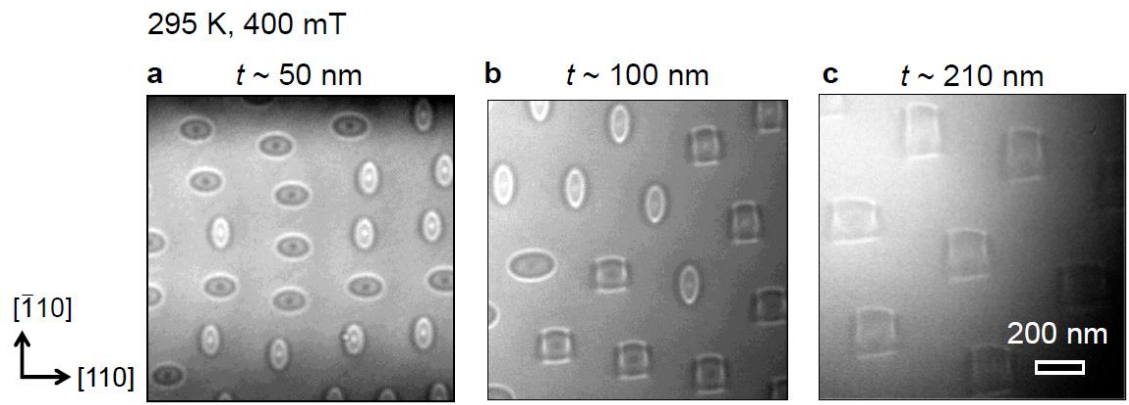

**Figure 3 | Lamella-thickness dependent transformation between skyrmions and antiskyrmions. a-c**, Under-focus LTEM images at 295 K and 400 mT taken in a similar protocol to Fig. 2a for lamella thickness of (a) 50 nm, (b) 100 nm and (c) 210 nm. The length scale is fixed for the three panels as indicated in the panel c.

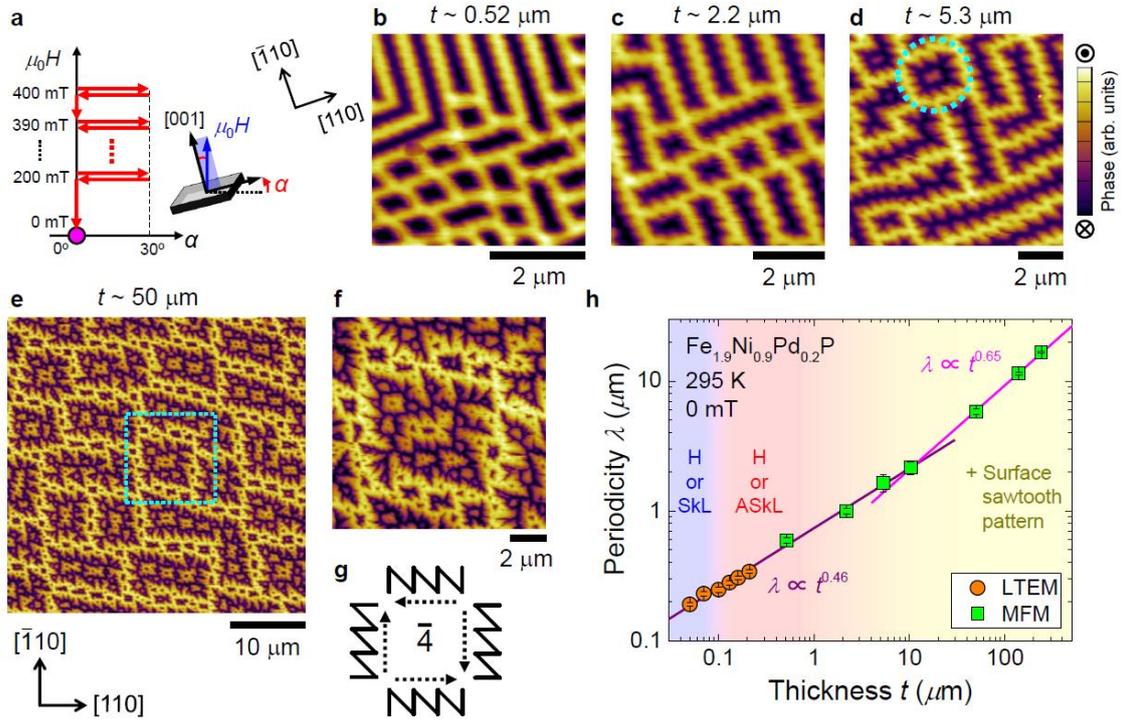

**Figure 4 | Thickness evolution of topological spin textures. a**, Schematic of the sample-tilting protocol under magnetic fields prior to the MFM measurements. A magnetic field of 400 mT is applied parallel to [001] direction, followed by tilting the sample plate to $\alpha \sim 30°$ and back to 0°. This procedure is repeated at different fields down to 200 mT (by 10 or 20 mT step) and then magnetic field is removed. MFM images (panels b-f) were taken at 295 K (±3 K) and 0 mT after this protocol. **b-f**, MFM images for the thickness of (b) 0.52 $\mu$m, (c) 2.2 $\mu$m, (d) 5.3 $\mu$m and (e, f) 50 $\mu$m, respectively. Here, the length scale is changed between each panel. Bright yellow and dark purple colors in the MFM images (as represented by the scale bar at right side of panel d) correspond to the magnetizations pointing out of the plane (parallel to [001]) and into the plane (antiparallel to [001]), respectively. A dotted circle in panel d indicates a domain with an inner core structure of reversed magnetization. The panel f corresponds to an enlarged view for the area indicated by a dotted square in the panel e. **g**, Schematic of sawtooth waves with $\bar{4}$ symmetry. The

dotted arrows indicate the direction of each sawtooth wave. All the MFM images at initial states including those for other thickness samples are presented in Supplementary Fig. 6. **h**, Crystal thickness ($t$) dependence of magnetic stripe periodicity ($\lambda$) at 295 K and 0 mT. Data points determined by LTEM and MFM measurements are denoted with orange circles and green squares, respectively. The error bar corresponds to the difference between maximum and minimum values observed within the same thickness region. The data points are fitted to a power law: $\lambda \sim t^{0.46}$ for $t \leq 10$ $\mu$m (purple line) and $\lambda \sim t^{0.65}$ for $t \geq 10$ $\mu$m (pink line). The blue and red shades in the background correspond to the thickness regions where Bloch-type skyrmions and antiskyrmions are observed, respectively, in Fig. 3. The yellow shade represents the thickness region where sawtooth patterns are found at the surface.

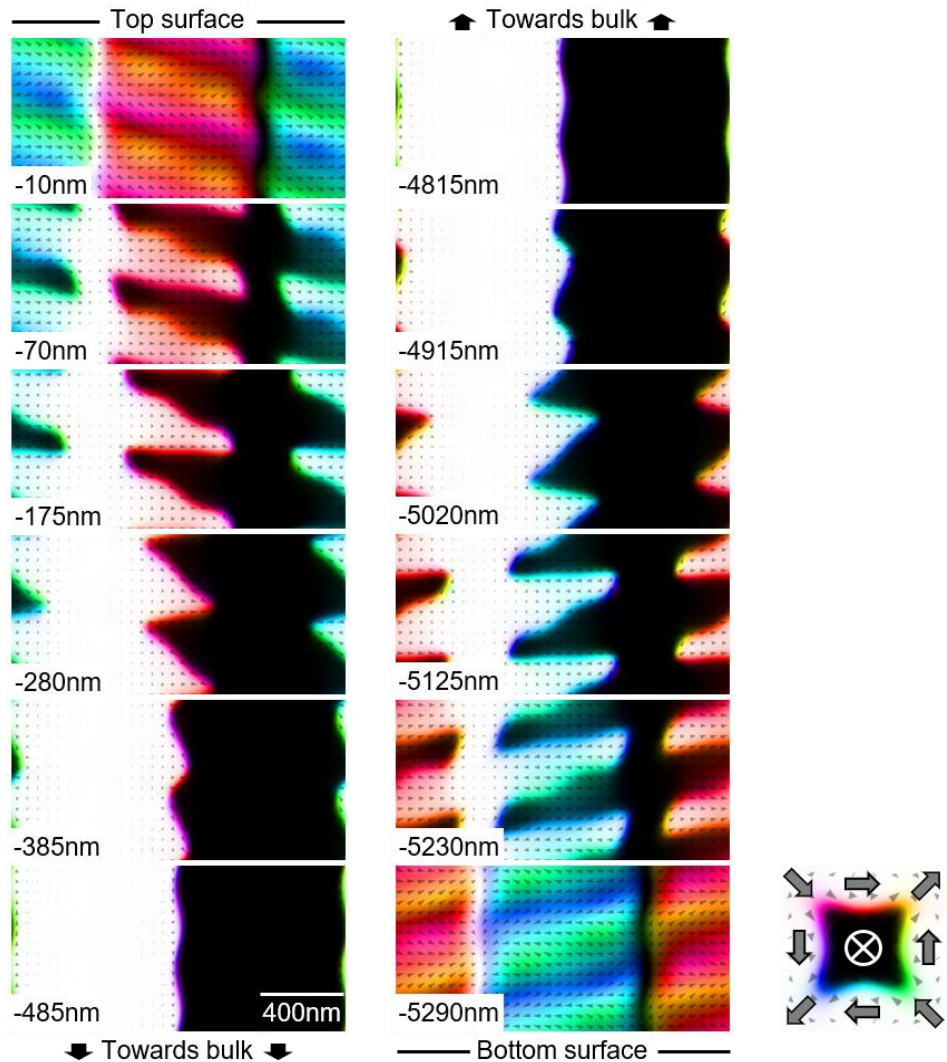

**Extended Data Fig. 1 | Micromagnetic simulations of the sawtooth magnetic texture.** The panels show the magnetization in various layers at different depth of a film as obtained from a three-dimensional micromagnetic simulation. The simulated sample measures 1.6 $\mu$m × 0.8 $\mu$m × 5.3 $\mu$m where periodic boundary conditions are applied in the x-y-plane to mimic an extended plate. The color encodes the direction of the magnetization in the plane and black/white encodes the out-of-plane component, as indicated by the square-shaped antiskyrmion on the bottom right panel which also sketches the DMI-preferred helicities. In addition, small arrows also show the direction of the in-plane components of the magnetization.

Supplementary information

**Room-temperature antiskyrmions and sawtooth surface textures in a noncentrosymmetric magnet with *S*$_4$ symmetry**

Kosuke Karube[1*], Licong Peng[1*], Jan Masell[1], Xiuzhen Yu[1], Fumitaka Kagawa[1,2], Yoshinori Tokura[1,2,3] & Yasujiro Taguchi[1]

*1. RIKEN Center for Emergent Matter Science (CEMS), Wako 351-0198, Japan.*

*2. Department of Applied Physics, University of Tokyo, Bunkyo-ku 113-8656, Japan.*

*3. Tokyo College University of Tokyo, Bunkyo-ku 113-8656, Japan.*

* These authors equally contributed

**Supplementary Note 1. Structural characterization**

Bulk single crystals of schreibersite (Fe, Ni)$_3$P and Pd-doped one (Fe, Ni, Pd)$_3$P were grown by a self-flux method from pure Fe, Ni and Pd metals and red phosphorous sealed in an evacuated quartz tube. Their chemical compositions were examined by energy dispersive X-ray (EDX) analysis for different sample lots, and summarized in Supplementary Tables 1 and 2. The results indicate that the ratio of the sum of transition metals to phosphorous is slightly smaller and larger than 3 for the samples without and with Pd, respectively, therefore we normalized the values so that the total amount of elements equals to 4. We cannot determine by the EDX results alone whether this is due to anti-site disorder between the metal and the phosphorous sites, or due to existence of vacancies at either site. Nevertheless, the analyzed compositions of the two compounds are approximately represented as Fe$_{1.6}$Ni$_{1.4}$P and Fe$_{1.9}$Ni$_{0.9}$Pd$_{0.2}$P, respectively, in average. We use these evaluated compositions throughout the main text and Supplementary information.

Phase purity in the bulk crystals of Fe$_{1.6}$Ni$_{1.4}$P and Fe$_{1.9}$Ni$_{0.9}$Pd$_{0.2}$P was confirmed by powder X-ray diffraction with Cu K$\alpha$ radiation as shown in Supplementary Figs. 1a and b, respectively. Rietveld refinements of the experimental data clearly show that both the samples crystalize in the $M_3$P-type noncentrosymmetric tetragonal structure with space group of $I\bar{4}$ (No. 82, $S_4^2$). This crystal structure contains three inequivalent transition-metal sites ($M$1, $M$2 and $M$3) and one P site, and 8 atoms for each site in a unit cell, as illustrated in the inset of Supplementary Fig. 1a. According to the previous structural study for (Fe, Ni)$_3$P in Ref. [1], Fe and Ni preferentially enter the $M$1 and $M$3 sites, respectively, while the $M$2 site is randomly occupied by both Fe and Ni. Therefore, we assume the site occupancies as $M$1 = Fe, $M$2 = Fe$_{0.6}$Ni$_{0.4}$ and $M$3 = Ni in the Rietveld analysis for $Fe_{1.6}Ni_{1.4}P$. From the refinement, tetragonal lattice parameters in $Fe_{1.6}Ni_{1.4}P$ are obtained as $a$ = 9.0262(6) Å and $c$ = 4.4565(3) Å ($c/a$ = 0.4937), showing good agreement with the previous studies [1, 2]. For the Rietveld refinement of $Fe_{1.9}Ni_{0.9}Pd_{0.2}P$, although the preferred site occupancy of a small amount of Pd is not uniquely determined by the Rietveld analysis based on the X-ray diffraction, we assume the site occupancies as $M$1 = Fe, $M$2 = $Fe_{0.9}Ni_{0.1}$ and $M$3 = $Ni_{0.8}Pd_{0.2}$, and obtain the good fitting result. The lattice parameters in $Fe_{1.9}Ni_{0.9}Pd_{0.2}P$ are $a$ = 9.1201(6) Å and $c$ = 4.4907(3) Å ($c/a$ = 0.4924), approximately 1% larger than those in $Fe_{1.6}Ni_{1.4}P$, evidencing that Ni is partially substituted by Pd with larger atomic radius.

The tetragonal $M_3P$ structure was also confirmed by Laue X-ray diffraction patterns for single crystals as shown in Supplementary Figs. 1c and d.

Therefore, we conclude that our target compound $Fe_{1.9}Ni_{0.9}Pd_{0.2}P$ has random site disorder among the three transition metals as well as slight anti-site disorder or vacancies, but the overall structure of this mixed crystal is of the $M_3P$ type with $S_4$ symmetry. As this magnet is an itinerant ferromagnet, the magnetic properties should be predominantly governed by the averaged structure and symmetry, instead of local ones.

**Supplementary Table 1 | Chemical composition of Pd-undoped samples determined by EDX**

| Lot No. | Fe | Ni | (Fe, Ni) | P |
|---|---|---|---|---|
| #1 | 1.63(1) | 1.34(1) | 2.97(0) | 1.03(0) |
| #2 | 1.61(1) | 1.36(2) | 2.96(1) | 1.04(1) |
| Average | 1.62(1) | 1.35(1) | 2.97(0) | 1.03(0) |

**Supplementary Table 2 | Chemical composition of Pd-doped samples determined by EDX**

| Lot No. | Fe | Ni | Pd | (Fe, Ni, Pd) | P |
|---|---|---|---|---|---|
| #1 | 1.93(1) | 0.85(1) | 0.24(1) | 3.02(2) | 0.98(2) |
| #2 | 1.90(1) | 0.89(1) | 0.26(1) | 3.05(1) | 0.95(1) |
| #3 | 1.95(2) | 0.88(1) | 0.21(1) | 3.04(1) | 0.96(1) |
| Average | 1.93(2) | 0.87(2) | 0.24(2) | 3.03(1) | 0.97(1) |

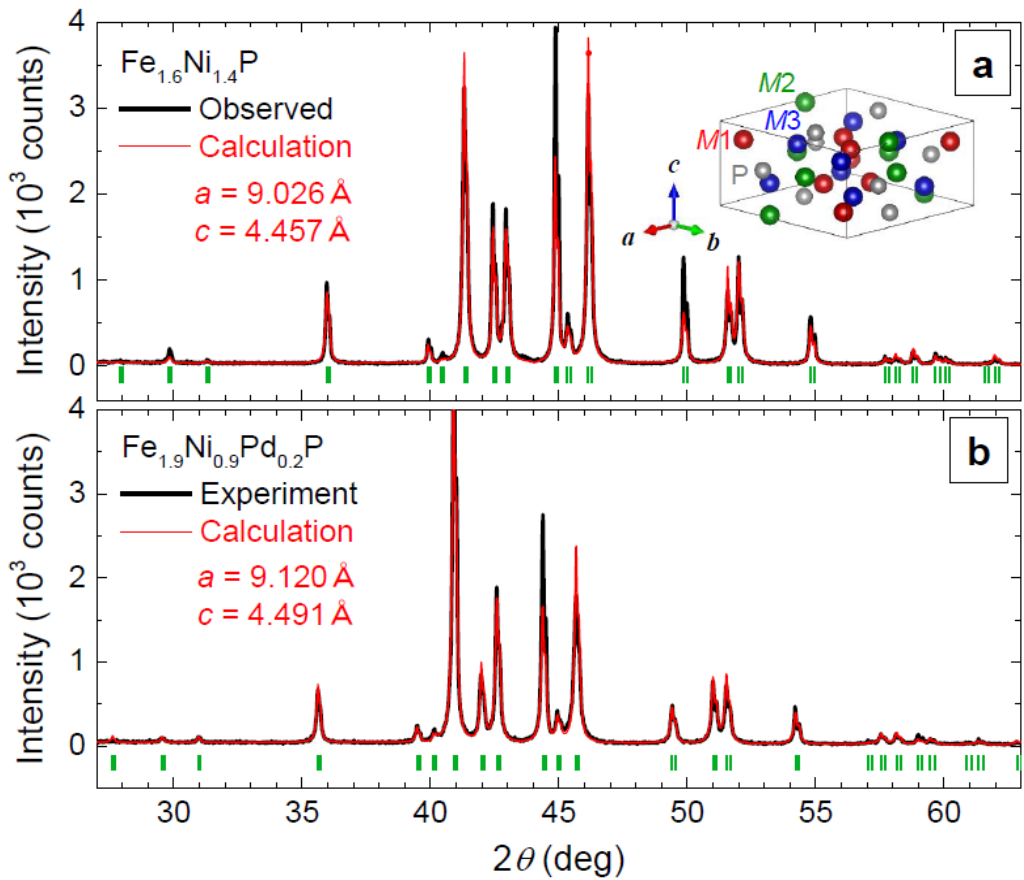

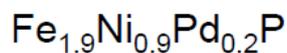

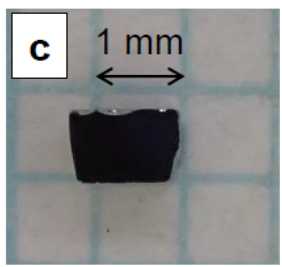

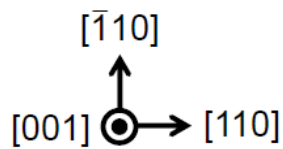

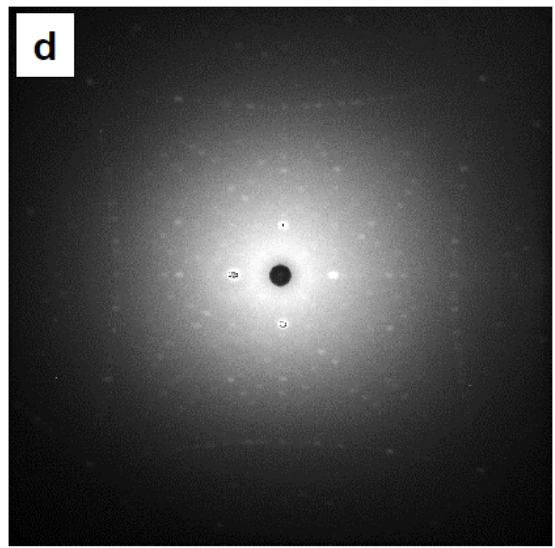

**Supplementary Figure 1 | Structural characterizations. a, b,** Powder X-ray diffraction profile at room temperature and Rietveld refinement for (a) $Fe_{1.6}Ni_{1.4}P$ and (b) $Fe_{1.9}Ni_{0.9}Pd_{0.2}P$. Experimental data are displayed with black line. Red line corresponds to

a simulation with a $M_3$P-type tetragonal structure as schematically shown in the inset of panel a. Peak positions are also indicated by green bars. **c,** Picture of a bulk single crystal of $Fe_{1.9}Ni_{0.9}Pd_{0.2}P$. **d,** Laue X-ray diffraction pattern of the (001) plane in the single crystal shown in panel c.

**Supplementary Note 2. Magnetic properties in schreibersite (Fe, Ni)$_3$P**

Magnetic properties for a single-crystalline bulk sample of Fe$_{1.6}$Ni$_{1.4}$P are presented in Supplementary Fig. 2. The observed ferromagnetic transition temperature ($T_c \sim 340$ K) and the saturation magnetization value at 2 K ($M_s \sim 3.2$ $\mu_B$/f.u.) agree with the previous studies for polycrystalline samples [2]. The magnetization curves show weak in-plane anisotropy, and there is no hysteresis in both directions. The area enclosed between the magnetization curves along the [001] direction (hard axis) and the [100] direction (easy axis) is estimated to be $\sim 33$ kJ/m$^3$. We have performed Lorentz transmission electron microscopy (LTEM) measurements for a (001) thin-plate of Fe$_{1.6}$Ni$_{1.4}$P but neither stripe patterns nor antiskyrmions were observed perhaps due to the in-plane anisotropy. Given this result, we partially substituted Ni with Pd, and found the change in anisotropy from in-plane to uniaxial one, in addition to probable enhancement of DMI, resulting in the emergence of stripe patters and antiskyrmions, as described in the main text.

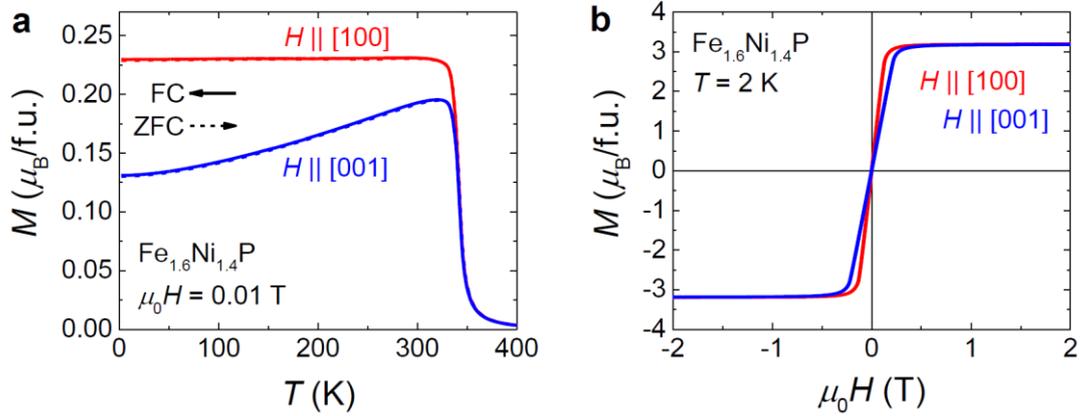

**Supplementary Figure 2 | Magnetizations in Fe$_{1.6}$Ni$_{1.4}$P. a**, Temperature ($T$) dependence of magnetization ($M$) under the magnetic field ($H$) of 0.01 T. $MT$ curves for $H \parallel [001]$ and $H \parallel [100]$ are presented with blue and red lines, respectively. The data taken during a field cooling (FC) and a warming after a zero-field cooling (ZFC) are denoted with solid and dotted lines, respectively. **b,** $MH$ curves at 2 K for $H \parallel [001]$ (blue line) and $H \parallel [100]$ (red line), indicating easy-plane-anisotropy feature.

**Supplementary Note 3. LTEM results in different measurement protocols**

We performed LTEM measurements for thin plates of $Fe_{1.9}Ni_{0.9}Pd_{0.2}P$ and found that magnetic states strongly depend on measurement processes, especially presence or absence of an in-plane component of the field. Supplementary Fig. 3 shows magnetic phase diagrams on the plane of temperature and magnetic field (with density mapping of the total number of closed magnetic objects superimposed) and representative LTEM images for different measurement protocols for a lamella (thickness: $t \sim 130$ nm). When the magnetic field applied perpendicular to the (001) plane is increased from an initial helical state (Supplementary Fig. 3a), only sparse antiskyrmions (and other magnetic objects) form in the background of helical or ferromagnetic states. This is probably because antiskyrmions are topologically disconnected from the helical state with a large energy barrier. This energy barrier can be overcome with an in-plane component of the field, which is applied by tilting the sample plate under magnetic fields (Supplementary Fig. 3b-d). As the field is increased while keeping the plate tilted by $\alpha \sim 12º$, a triangular lattice of dense magnetic bubbles forms (Supplementary Fig. 3b). This magnetic bubble is topologically trivial and thus termed "non-topological (NT-) bubble" as in Ref. [3], which is composed of a square half-antiskyrmion and a circular or elliptical half-skyrmion, resulting in the D- or bullet-like shape. As also described in the main text (Fig. 2), after creating the dense NT-bubbles, the sample plate was tilted back to the initial orientation ($\alpha \sim 0º$), then the NT-bubbles transform to dense square antiskyrmions forming a lattice state (Supplementary Fig. 3c). In the subsequent field increasing process with $\alpha \sim 0º$, the antiskyrmions persist and finally transform to elliptical skyrmions with different helicities at high fields, via a coexistence state with NT-bubbles. Antiskyrmions that are once created by the tilting procedure persist down to zero field as a metastable state although some of them are relaxed to helical stripes (Supplementary Fig. 3d). Subsequently, as the field is increased again while keeping $\alpha \sim 0º$, the metastable antiskyrmions persist over a wide field region and finally transform to NT-bubbles and skyrmions at high fields (Supplementary Fig. 3d). While various magnetic phase diagrams depending on tilting procedures are similar to those reported in $Mn_{1.4}Pt_{0.9}Pd_{0.1}Sn$ [3], the normal-field-induced topological transition from the antiskyrmion lattice to the skyrmion lattice observed over a wide temperature region is unique to the present compound.

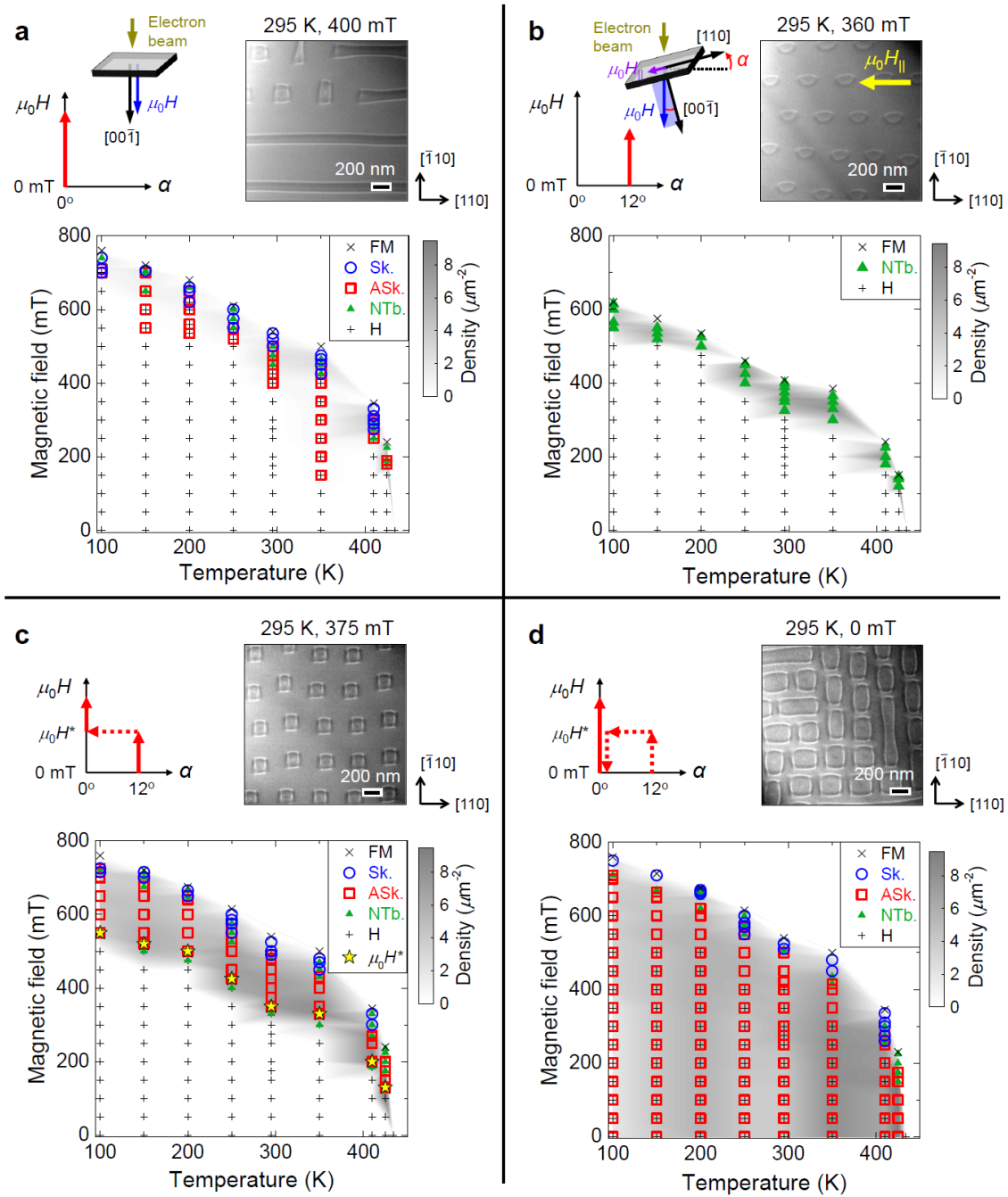

**Supplementary Figure 3 | LTEM results in different measurement protocols for a lamella with a thickness ($t \sim 130$ nm). a-d**, Under-focus LTEM images and magnetic phase diagrams on the plane of temperature and magnetic field ($\mu_0 H$) in four different measurement processes as schematically shown in each panel. Here, $\alpha$ is defined as the tilt angle of the sample plate from the normal plane to the electron beam and the field

(inset of panel b), equivalent to the tilt angle of the field from the $[00\bar{1}]$ axis. (a) The field is increased without tilting the sample plate. (b) The field is increased while the sample plate is tilted by $\alpha \sim 12°$. (c) After the field is increased with the tilt of $\alpha \sim 12°$, the sample plate is tilted back to $\alpha \sim 0°$ at $\mu_0 H^*$, and the field is increased again. (d) After the tilting procedure in panel c, once the field is decreased down to zero, and subsequently the field is increased without tilting. In the phase diagrams, the observed magnetic textures [helical structure (H), non-topological bubbles (NTb.), antiskyrmions (ASk.), skyrmions (Sk.) and ferromagnetic structure (FM)] at each temperature and magnetic-field point are denoted with different symbols. The total density of all the closed magnetic objects (non-topological bubbles, antiskyrmions and skyrmions) is also plotted as the gray color in the state diagrams. The color scale is fixed for all the panels.

**Supplementary Note 4. Field-induced transformation from antiskyrmions to skyrmions**

The detail of transformation from antiskyrmions to skyrmions upon increasing the magnetic field at 295 K during the process of Supplementary Fig. 3c is shown in Supplementary Video 1: square-shaped antiskyrmions form a triangular lattice in the field range of 375 − 470 mT; square-shaped antiskyrmions transform to elliptical skyrmions through a coexistence state with bullet-shaped NT-bubbles in fields of 475 − 495 mT; elliptical skyrmions with random helicities remain in fields of 500 − 530 mT; a monotonic contrast (a ferromagnetic state) shows up in stronger fields above 535 mT.

**Supplementary Note 5. DPC-STEM observation of antiskyrmions**

In the main text (Fig, 2), we present magnetic configurations of square-shaped antiskyrmions with mapping in-plane magnetic induction fields (*B*-fields) by analyses of defocused LTEM images using transport-of-intensity equation (TIE) [4]. The antiskyrmion has four Bloch lines (Néel-type domain walls inserted in Bloch-type domain walls with opposite helicities) which produce stray fields to neighboring antiskyrmions.

To quantitatively map the local *B*-field and characterize Bloch lines in antiskyrmions, we use an in-focus magnetic imaging technique, the differential phase contrast (DPC) imaging at scanning transmission electron microscopy (STEM) mode [5−11]. In Supplementary Fig. 4, we compare in-plane *B*-field maps of antiskyrmions obtained by TIE analyses of defocused LTEM images (Supplementary Fig. 4c, d) and by DPC-STEM (Supplementary Fig. 4g, h), which were acquired at the same sample area at zero field. The *B*-field obtained by the two techniques are almost identical. Therefore, the TIE image precisely represents the in-plane *B*-fields. The line profiles of $B_x$ and $B_y$ components extracted from the DPC-STEM measurements clearly show that the Bloch-type walls are parallel to the [110] axis while the Bloch lines (Néel-type walls) are parallel to the [100] axis. We quantitatively estimate the width of Bloch-type domain walls in the antiskyrmion to be ~ 52 nm (full width at half maximum ~ 26 nm).

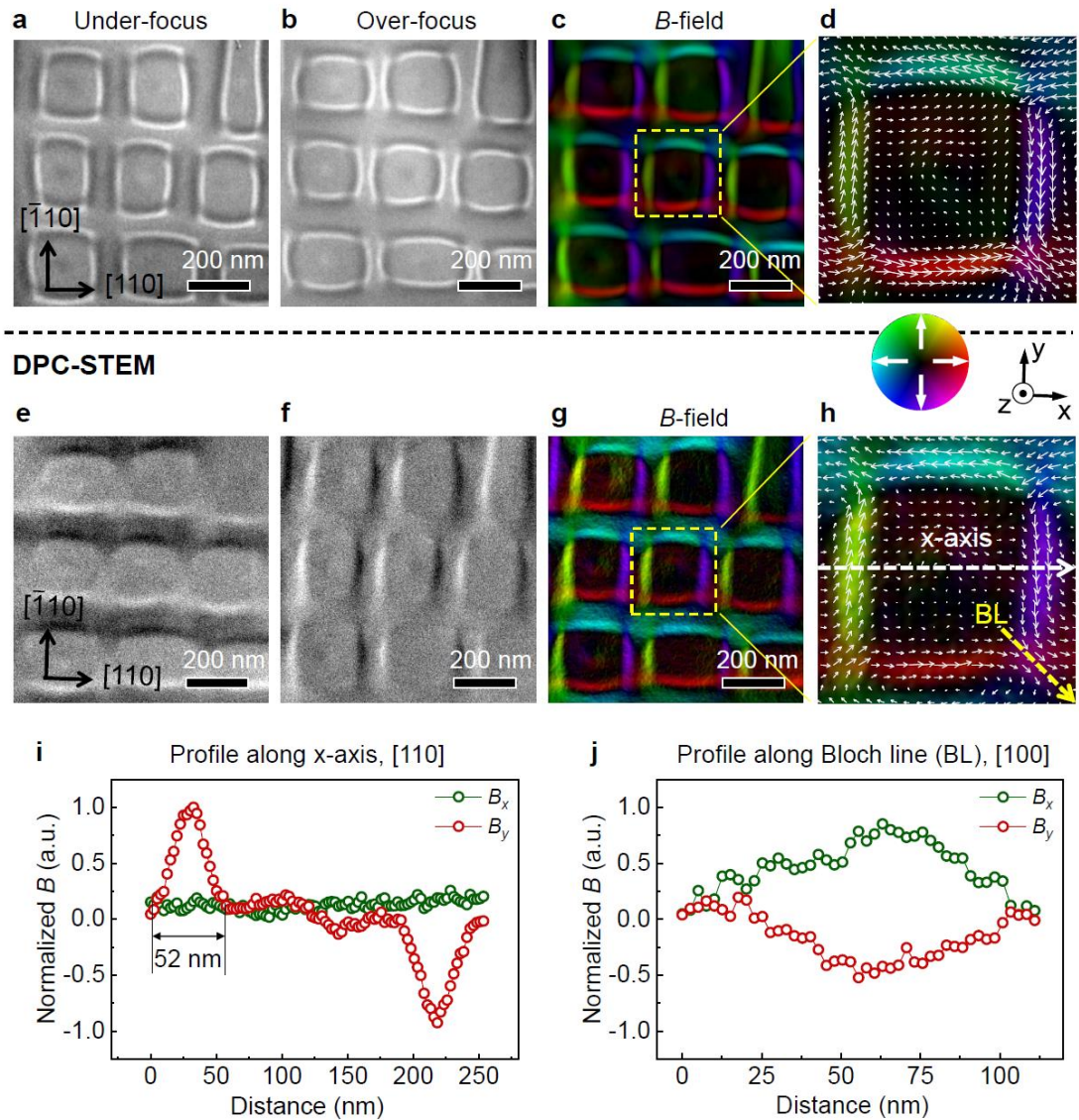

**Supplementary Figure 4 | Magnetic induction field maps of antiskyrmions obtained by TIE analyses and DPC-STEM. a-d,** TIE analyses of (c, d) magnetic induction field of antiskyrmions from (a) under-focus and (b) over-focus LTEM Fresnel images with ∓200 μm defocus. **e-j,** DPC-STEM images of antiskyrmions at in-focus condition obtained at the same area with the LTEM imaging: (e, f) DPC images along x- and y-direction, respectively, (g, h) the *B*-field map, and (i, j) profiles along x-axis and along the Bloch line (BL) (as marked with white and yellow lines) showing the normalized $B_x$,

$B_y$ components along x- and y-axes. The domain wall width is estimated to be ~ 52 nm as indicated in the x-axis profile. All the data are obtained at room temperature and zero field, following the same experimental process as Supplementary Fig. 3.

**Supplementary Note 6. LTEM observations of thickness-dependent stripe patterns**

To discuss the dependence of the periodicity of magnetic helices on crystal thickness, we present LTEM observations of initial helical states at 295 K and zero field for a lamella with various thicknesses ranging from 50 nm to 210 nm in Supplementary Fig. 5. As shown in the periodicity vs. thickness plot (Supplementary Fig. 5b) based on LTEM images of helical stripes (Supplementary Fig. 5c-h), the magnetic periodicity monotonically increases as the thickness is increased. The data points in Supplementary Fig. 5b are also plotted in logarithmic scale in Fig. 4h in the main text, together with magnetic force microscopy (MFM) data.

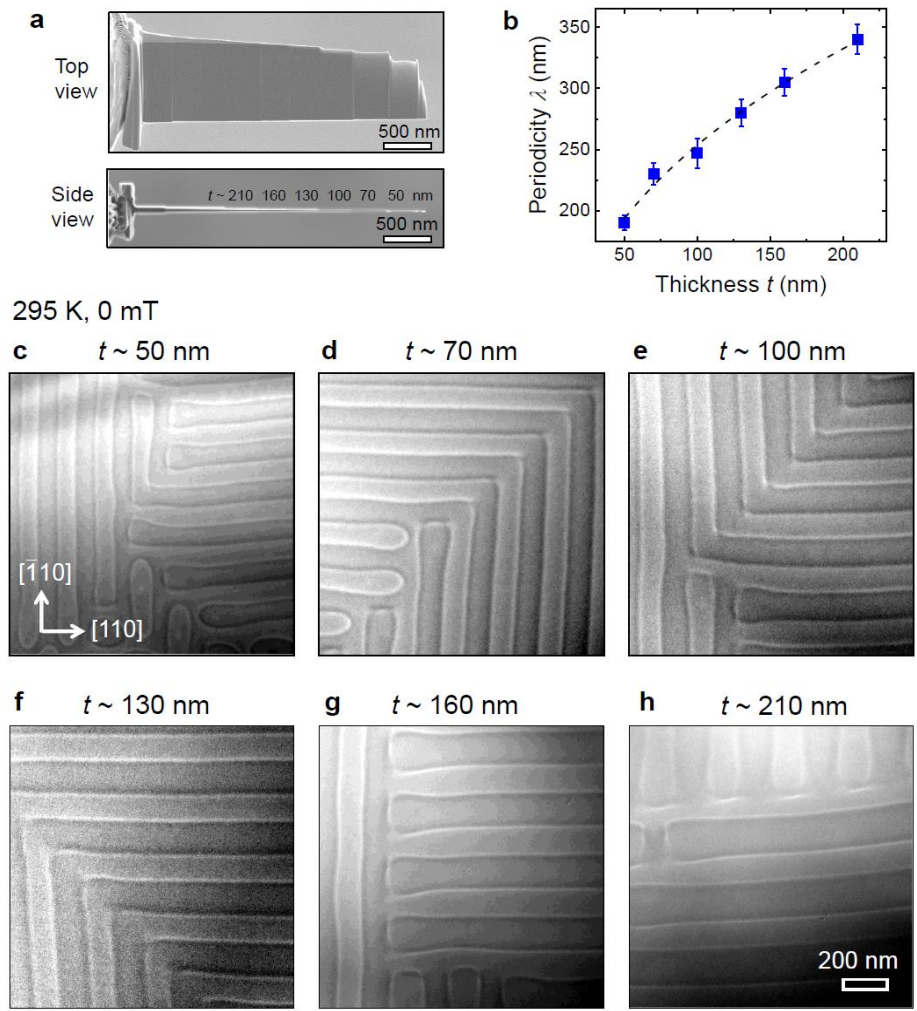

**Supplementary Figure 5 | LTEM results of magnetic stripe patterns in a lamella with various thicknesses. a,** Scanning electron microscopy (SEM) images of the top and side views of a staircase-like thin plate with various thicknesses ranging from 50 nm to 210 nm. **b,** Magnetic stripe periodicity ($\lambda$) as a function of lamella thickness ($t$) obtained from the LTEM images in panels c-h. **c-h**, Under-focus LTEM images at 295 K and 0 mT for the lamella thickness of (c) 50 nm, (d) 70 nm, (e) 100 nm, (f) 130 nm, (g) 160 nm and (h) 210 nm. The length scale is fixed for the 6 panels as indicated in the panel h.

**Supplementary Note 7. MFM observations of thickness-dependent surface magnetic textures**

To discuss the thickness-dependent surface magnetic textures in thick crystals, we present MFM images in Supplementary Fig. 6 that are measured at 295 K and zero field before applying magnetic fields for crystals with various thicknesses ranging from $t \sim$ 0.52 $\mu$m to 240 $\mu$m. For $t \sim 0.52$ $\mu$m, a stripe pattern propagating along [110] direction is observed (Supplementary Fig. 6a). Unlike the straight stripes observed by LTEM, a weak undulation in the domain walls is discernable. The wavy domain walls are more clearly observed for $t \sim 2.2$ $\mu$m (Supplementary Fig. 6b). The domain walls become rather sharp sawtooth-like waves for $t \sim 5.3$ $\mu$m, where smaller dot-like subdomains with reversed magnetization are barely observed (Supplementary Fig. 6c). The number of the subdomains inside the main sawtooth stripes increases for $t \sim 10.3$ $\mu$m (Supplementary Fig. 6d). These sawtooth domain walls and subdomains become more complex for $t \sim 50$ $\mu$m (Supplementary Fig. 6e and f), exhibiting fractal patterns, i.e. self-similar objects in multiple length scales, like the Sierpinski carpet [12]. For thicker crystals with $t \sim 140$ $\mu$m (Supplementary Fig. 6g and h) and 240 $\mu$m (Supplementary Fig. 6i and j), the sawtooth domains are further branched and even more complex fractal patterns are observed. Importantly, the sawtooth patterns in the vertical stripe with $q \parallel [110]$ and horizontal stripe with $q \parallel [\bar{1}10]$ should be reversed in order for them to coincide with each other; namely, they are converted by $\bar{4}$ symmetry operation, thus characteristic for this antiskyrmion material with $S_4$ symmetry. The magnetic periodicity ($\lambda$) of the main sawtooth stripes increases with the crystal thickness ($t$) from $\lambda \sim 0.6$ $\mu$m ($t \sim 0.52$ $\mu$m) up to $\lambda \sim 17$ $\mu$m ($t \sim 240$ $\mu$m) as plotted in Fig. 4h in the main text.

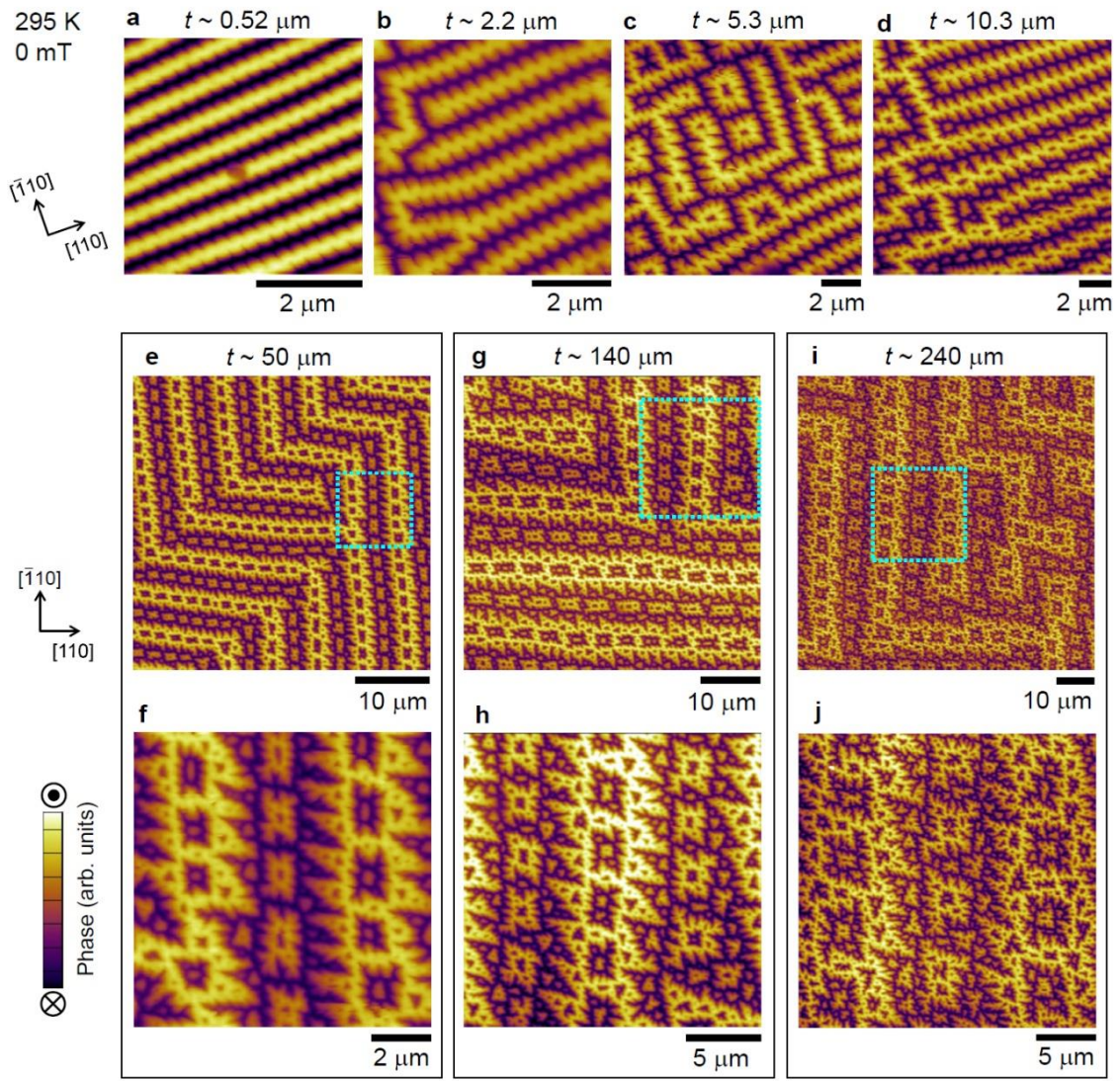

**Supplementary Figure 6 | MFM results of surface magnetic stripe patterns in various crystal thicknesses. a-j,** MFM images measured at room temperature (295 K ± 3 K) and zero field before applying magnetic fields (or after a zero-field cooling from 400 K) for the thickness of (a) 0.52 $\mu$m, (b) 2.2 $\mu$m, (c) 5.3 $\mu$m, (d) 10.3 $\mu$m, (e, f) 50 $\mu$m, (g, h) 140 $\mu$m and (i, j) 240 $\mu$m. The panels f, h and j correspond to enlarged MFM images for the areas indicated by dotted squares in the panels e, g and i, respectively. The length scale is changed between each panel.

**Supplementary Note 8. Interpretation of thickness dependence of magnetic textures**

The dependence of magnetic domain structures on the crystal thickness in uniaxial ferromagnets has been studied for a long time. Here, we interpret our MFM results on the basis of the established theoretical models for uniaxial magnets, following the discussion in Ref. [13]. The domain structure in sufficiently thin films with large uniaxial anisotropy is well described by the Kittel's parallel-plate domain model. In this model, by using a saturation magnetization $M_s$, an effective uniaxial anisotropy constant $K_u$ and a domain wall energy $\gamma$, the dependence of the domain periodicity $\lambda$ on the film thickness $t$ is given by

$$\lambda = \left[\frac{\gamma(1+\mu^{*1/2})}{3.4M_s^2}\right]^{1/2} t^{1/2}.$$

Here, $\mu^*$ is a rotational permeability defined as $\mu^* = 1 + Q^{-1} = 1 + 2\pi M_s^2/K_u$, which was introduced to describe deviation of magnetizations from the easy axis due to demagnetization ($\mu^* = 1$ in the Kittel's original model [14]). As the thickness is increased, the domain structure near the surface becomes more complex to reduce magnetostatic energy. Above the critical thickness $t_{c1}$, domain walls undulate near the surface. The theoretical model of this wavy domain structure by Szymczak [15] shows the relation

$$\lambda = 0.54\gamma^{0.4}\mu^{*0.2}M_s^{-0.8}t^{0.6} \left(t_{c1} = 26.39\frac{\gamma}{\mu^{*2}M_s^2}\right).$$

Above the critical thickness $t_{c2}$, the main domains are further branched and reversely magnetized spike structures appear at the surface. This surface spike structure was theoretically considered by Kaczér [16], where the periodicity of the main domains is given by

$$\lambda = \left(\frac{3^2\gamma\mu^*}{8^2\pi M_s^2}\right)^{1/3} t^{2/3} \left(t_{c2} = \frac{16\pi^2\gamma}{1.7^3\mu^{*2}M_s^2}\right).$$

These theoretical models were experimentally confirmed in many uniaxial ferromagnets, for example, in a hexagonal ferrite $PbFe_{12}O_{19}$ ($Q$ = 3.4, $\mu^*$ = 1.3) [16].

As presented in Fig. 4h in the main text, in our LTEM and MFM experiments for $Fe_{1.9}Ni_{0.9}Pd_{0.2}P$, we found power-law relations as $\lambda \sim t^{0.46}$ ($t \leq 10$ $\mu$m) and $\lambda \sim t^{0.65}$ ($t \geq 10$ $\mu$m), in which the exponents are very close to those in the Kittel's model and the Kaczér's one, respectively. If we assume that the prefactor of $\lambda(t)$ for $t \leq 10$ $\mu$m equals to the Kittel's formula, by using the values obtained from magnetizations at 300 K ($M_s$ = 417 emu/cm$^3$, $K_u$ = 3.1 × 10$^5$ erg/cm$^3$, $Q$ = 0.28, $\mu^*$ = 4.6), the domain wall energy is obtained as $\gamma$ = 5.0 erg/cm$^2$. From this value, the critical thickness in the Szymczak's model and the Kaczér's one is calculated as $t_{c1}$ = 0.36 $\mu$m and $t_{c2}$ = 0.44 $\mu$m, respectively. In our MFM experiments, wavy domain walls are observed already at $t \sim 0.52$ $\mu$m and more clearly above $t \sim 2.2$ $\mu$m. Therefore, the calculated critical thickness shows a good agreement with the experimental result. We should note that the presence of the Dzyaloshinskii-Moriya interaction (DMI) in $Fe_{1.9}Ni_{0.9}Pd_{0.2}P$ may quantitatively alter the results of these models. In particular, the DMI contributes as a linear term to the domain wall energy as often described by

$$\gamma = 4\sqrt{AK_u} - \pi D,$$

where $A$ is the exchange stiffness and $D$ is the DMI constant [17]. Nevertheless, since the anisotropic DMI in the present system mainly contributes to the anisotropic feature of the domain wall as detailed in the following micromagnetic simulations, the main feature of the thickness dependence, which is determined by the competition between the gain in the magnetostatic energy and the cost in the domain wall energy, can be captured.

**Supplementary Note 9. Micromagnetic simulations for sawtooth patterns**

In order to better understand the origin of the sawtooth pattern we have performed micromagnetic simulations, using the well-established GPU-accelerated program MuMax3 [18]. We personalized the code to incorporate the asymmetric DMI, such that the energy functional now reads

$$E[m] = \int_V A (\nabla m)^2 + D [m \cdot (\hat{x} \times \partial_x m) - m \cdot (\hat{y} \times \partial_y m)] - K_u (m \cdot \hat{z})^2 - \frac{M_s}{2} m \cdot B_d \, dV$$

where $B_d$ is the demagnetizing field. We use the experimentally obtained values at a temperature of $T = 300$ K for the uniaxial anisotropy $K_u = 31$ kJ/m$^3$ and the magnetization $M_s = 417$ kA/m. The magnetic stiffness at $T = 0$ K is estimated via the critical temperature $T_c = 400$ K as $A_0 = \frac{\left(\frac{a}{2}\right)^2 T_c M_s}{2 g \mu_B} = 20.6$ pJ/m, where we used $g = 2$ and the lattice constant $a = 9.1215$ Å. Moreover, we account for the finite temperature by correcting the stiffness as $A(300 \text{ K}) = A_0 \left(\frac{M_s(300 \text{ K})}{M_s(0 \text{ K})}\right)^2 = 8.1$ pJ/m. The resulting length scale for domain walls at this temperature is therefore $\sqrt{A/K_u} = 16$ nm (full width at half maximum $\sim 2.6\sqrt{A/K_u} = 42$ nm), which agrees with the experimental observations. We treat the DMI as a small perturbation and checked various values $D = 0, 0.05, 0.1, 0.2$ mJ/m$^2$. Moreover, we improved the discretization scheme for the derivatives in the continuum model by using fourth order finite difference stencils as described in Ref. [19]. This modification allows us to use smaller lattices while keeping the numerical error low. The result for a 1.6 μm × 0.8 μm × 5.3 μm large system with periodic boundary conditions in the x-y-plane (discretized on 512 × 256 × 256 lattice sites) is presented in Extended Data Fig. 1 in the main text for $D = 0.2$ mJ/m$^2$. The periodicity is adapted from the experiments. The sawtooth patterns are clearly observed near the both surfaces while the domain wall gradually changes to straight line with very tiny modulation as the distance from the surface is increased deep into the bulk. We checked that the sawtooth pattern also appears without DMI, but the DMI enhances the effect. In the following, we give a qualitative explanation for this effect, which is mainly driven by the unique handedness of domain walls and the competition between the magnetic stiffness and demagnetization energy.

**Supplementary Note 10. Qualitative explanation of sawtooth patterns**

Given the above simulation results, we qualitatively discuss the origin of the sawtooth pattern with the schematic figures in Supplementary Figs. 7a-g. In the bulk of the thick crystal, the domain walls prefer to be Bloch type to minimize magnetic volume charges, and their helicity is fixed by the anisotropic DMI as clockwise (CW) and counterclockwise (CCW) manners along the two orthogonal [110] and [$\bar{1}$10] axes, respectively (Supplementary Figs. 7d, e). Near the surfaces, on the other hand, the domain wall structures must be modified in order to reduce surface magnetic charges while the transformation from the bulk pattern to the surface pattern should be continuous. The longer parts of the domain wall are smoothly connected to the bulk state and they slightly tilt due to the stray field from the domains on the surface. The shorter parts of the domain wall reverse their helicity and thereby nearly perfectly match the stray field direction. Importantly, the reversed helicity of the shorter parts is also preferred by the DMI. To change the helicity, however, the magnetic moments have to rotate near the surface, which locally costs energy due to the magnetic stiffness, but this is only a one-dimensional defect while the domain wall with the original helicity is two-dimensional. Moreover, the sharp edges of the zigzag sawtooth pattern have magnetic charges but this is also a small fraction of the entire domain wall. Totally, the sawtooth pattern can have lower energy due to the large energy gain in the demagnetization and the DMI at the small cost of the domain wall energy.

Reflecting the anisotropic feature of the DMI-induced helicity in the bulk, the sawtooth pattern arising from the vertical stripes [Supplementary Fig. 7b(f)] and that from the horizontal ones [Supplementary Fig. 7c(g)] coincide with each other upon 90º rotation followed by inversion ($\bar{4}$). Since the stray field direction is reversed for the top and bottom surfaces, the tilt directions of the domain walls are opposite on the two surfaces, and consequently the sawtooth patterns for the top and the bottom surfaces are reversed when viewed from the same direction.

To confirm the above explanation, we present MFM images on both sides of a thick crystal ($t \sim 240$ $\mu$m) in Supplementary Figs. 7h [front side, (001) plane] and i [back side, (00$\bar{1}$) plane]. The observed sawtooth directions are unchanged for both sides. Therefore, as schematically illustrated in Supplementary Fig. 7j, the sawtooth pattern must be reversed for the top surface and for the bottom surface when viewed from +$z$ side of the sample, in accordance with the micromagnetic simulation and the above qualitative discussion.

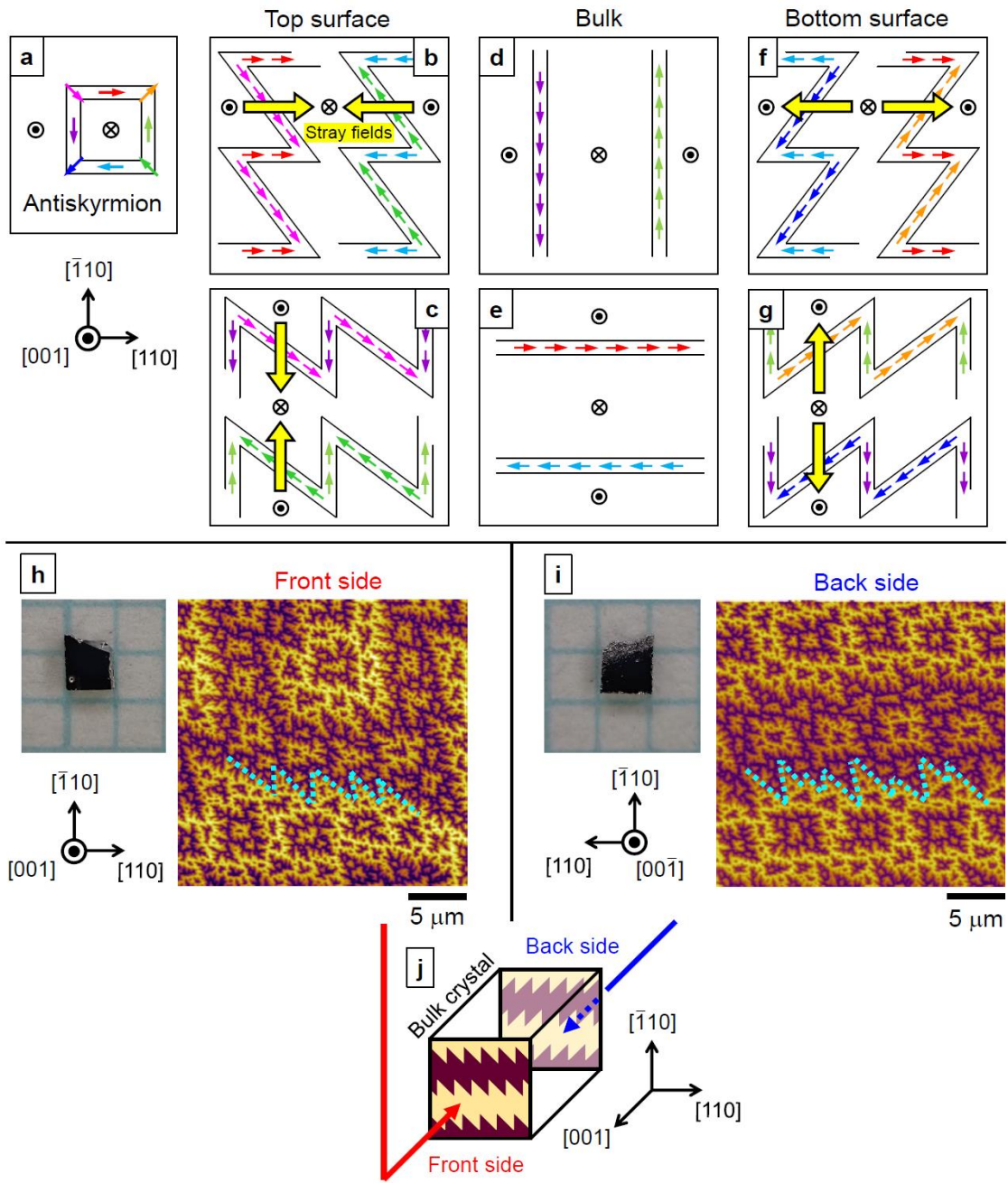

**Supplementary Figure 7 | Qualitative explanation of the sawtooth pattern. a**, Schematic of magnetic moments of an antiskyrmion in the present system favored by the anisotropic DMI. **b-g**, Schematics of magnetic domain walls with $q \parallel [110]$ (top panels) and $q \parallel [\bar{1}10]$ (bottom panels) induced by the DMI and dipolar interaction (b, c) near the top surface, (d, e) in the bulk, and (f, g) near the bottom surface. All the figures are viewed

from +*z* side of the plate. The colors of the small arrows indicate the directions of in-plane magnetic moments. The direction of the stray fields between the domains is shown by thick yellow arrows. **h, i,** Sample photos and MFM images on (h) the front side [(001) plane] and (i) the back side [(00$\bar{1}$) plane] of a bulk crystal with a thickness of *t* ~ 240 *μ*m. The back-side image is obtained after the sample is rotated by 180º around the [$\bar{1}$10] axis. The characteristic sawtooth domain walls are highlighted by dotted light-blue lines. **j,** Schematic sawtooth domain patterns at the two surfaces of the bulk crystal. The sawtooth patterns on the front and back sides correspond to the observed MFM images in panels h and i, respectively.

**Supplementary References**

1.  Moretzki, O. *et al*. Determination of the metal ordering in meteoritic (Fe,Ni)$_3$P crystals. *J. Synch. Rad.* **12**, 234–240 (2005).

2.  Goto, M., Tange, H., Tokunaga, T., Fujii, H. & Okamoto, T. Magnetic Properties of the (Fe$_{1-x}$M$_x$)$_3$P Compounds. *Jpn. J. Appl. Phys.* **16**, 2175−2179 (1977).

3.  Peng, L. C. *et al*. Controlled transformation of skyrmions and antiskyrmions in a non-centrosymmetric magnet. *Nat. Nanotech.* **15**, 181–186 (2020).

4.  Ishizuka, K. & Allman, B. Phase measurement of atomic resolution image using transport of intensity equation. *J. Electron Microsc.* **54**, 191–197 (2005).

5.  Chapman, J. N., Batson, P. E., Waddell, E. M. & Ferrier, R. P. The direct determination of magnetic domain wall profiles by differential phase contrast electron microscopy. *Ultramicroscopy* **3**, 203–214 (1978).

6.  Sandweg, C. W. *et al*. Direct observation of domain wall structures in curved permalloy wires containing an antinotch. *J. Appl. Phys.* **103**, 093906 (2008).

7.  McGrouther, D. *et al*. Internal structure of hexagonal skyrmion lattices in cubic helimagnets. *New J. Phys.* **18**, 095004 (2016).

8.  Shibata, N. *et al*. Direct Visualization of Local Electromagnetic Field Structures by Scanning Transmission Electron Microscopy. *Acc. Chem. Res.* **50**, 1502−1512 (2017).

9.  Matsumoto, T., So, Y. G., Kohno, Y., Ikuhara, Y. & Shibata, N. Stable Magnetic Skyrmion States at Room Temperature Confined to Corrals of Artificial Surface Pits Fabricated by a Focused Electron Beam. *Nano Lett.* **18**, 754–762 (2018).

10. Pöllath, S. *et al*. Spin structure relation to phase contrast imaging of isolated magnetic Bloch and Néel skyrmions. *Ultramicroscopy* **212**, 112973 (2020).

11. Yasin, F. S. *et al*. Bloch lines constituting antiskyrmions captured via differential phase contrast. *Adv. Mater.* **32**, 2004206 (2020).

12. Sierpinski, W. Sur une courbe cantorienne qui contient une image biunivoque et continue de toute courbe donnée. *C. R. Acad. Sci. Paris* **162**, 629–632 (1916).

13. Szmaja, W. Investigations of the domain structure of anisotropic sintered Nd–Fe–B-based permanent magnets. *J. Mag. Mag. Mater.* **301** 546–561 (2006).

14. Kittel, C. Theory of the Structure of Ferromagnetic Domains in Films and Small Particles. *Phys. Rev.* **70**, 965–971 (1946).

15. Szymczak, R. A Modification of the Kittel Open Structure. *J. Appl. Phys.* **39**, 875–876 (1968).

16. Kaczér, J. On the domain structure of uniaxial ferromagnets. *Sov. Phys. JETP* **19**, 1204–1208 (1964).

17. Heide, M., Bihlmayer, G. & Blügel, S. Dzyaloshinskii-Moriya interaction accounting for the orientation of magnetic domains in ultrathin films: Fe/W(110). *Phys. Rev. B* **78**, 140403(R) (2008).

18. Vansteenkiste, A. *et al*. The design and verification of MuMax3. *AIP Adv.* **4**, 107133 (2014).

19. Müller, J. *Magnetic Skyrmions and Topological Domain Walls*. Universität zu Köln (2018). https://kups.ub.uni-koeln.de/8140